\documentclass[10pt,aps,reprint,prb,floatfix,superscriptaddress,longbibliography]{revtex4-2}
\pdfoutput=1
\usepackage{graphicx,latexsym}
\usepackage{dcolumn,amsfonts}
\usepackage{amssymb,amsmath,bm}

\usepackage[breaklinks=true,pdfencoding=auto]{hyperref}

\usepackage{natbib}
\usepackage{balance}

\hypersetup{
	colorlinks   = true, %Colours links instead of ugly boxes
	urlcolor     = blue, %Colour for external hyperlinks
	linkcolor    = blue, %Colour of internal links
	citecolor    = red %Colour of citations
}

\usepackage{color}
\usepackage{ulem}
\begin{document}

%-----------------------------------------------------------------

\title{Driven square lattice of quantum dots in a magnetic field\\ coupled to a cylindrical FIR-photon cavity}

\author{Vidar Gudmundsson}
\email{vidar@hi.is}
\affiliation{Science Institute, University of Iceland, Dunhaga 3, IS-107 Reykjavik, Iceland}
\affiliation{Department of Engineering, Reykjavik University, Menntavegur 1, IS-102 Reykjavik, Iceland}
\author{Vram Mughnetsyan}
\email{vram@ysu.am}
\affiliation{Department of Condensed Matter Physics, Yerevan State University, Alex Manoogian 1, 0025 Yerevan, Armenia}
\author{Hsi-Sheng Goan}
\email{goan@phys.ntu.edu.tw}
\affiliation{Department of Physics and Center for Theoretical Physics, National Taiwan University, Taipei 106319, Taiwan}
\affiliation{Center for Quantum Science and Engineering, National Taiwan University, Taipei 106319, Taiwan}
\affiliation{Physics Division, National Center for Theoretical Sciences, Taipei 106319, Taiwan}
\author{Jeng-Da Chai}
\email{jdchai@phys.ntu.edu.tw}
\affiliation{Department of Physics and Center for Theoretical Physics, National Taiwan University, Taipei 106319, Taiwan}
\affiliation{Center for Quantum Science and Engineering, National Taiwan University, Taipei 106319, Taiwan}
\affiliation{Physics Division, National Center for Theoretical Sciences, Taipei 106319, Taiwan}
\author{Nzar Rauf Abdullah}
\email{nzar.r.abdullah@gmail.com}
\affiliation{Physics Department, College of Science, University of Sulaimani, Kurdistan Region, Iraq}
\author{Chi-Shung Tang}
\email{cstang@nuu.edu.tw}
\affiliation{Department of Mechanical Engineering, National United University, Miaoli 360302, Taiwan}
\author{Wen-Hsuan Kuan}
\email{wenhsuan.kuan@gmail.com}
\affiliation{Department of Applied Physics and Chemistry, University of Taipei, Taipei 10048, Taiwan}
\affiliation{Science Institute, University of Iceland, Dunhaga 3, IS-107 Reykjavik, Iceland}
\author{Valeriu Moldoveanu}
\email{valim@infim.ro}
\affiliation{National Institute of Materials Physics, PO Box MG-7, Bucharest-Magurele, Romania}
\author{Andrei Manolescu}
\email{manoles@ru.is}
\affiliation{Department of Engineering, Reykjavik University, Menntavegur 1, IS-102 Reykjavik, Iceland}

%
%----------------------------------------------------------------

\begin{abstract}
We present a comprehensive computational study of driven quantum dot arrays in a square lattice configuration, subject to an external magnetic field and coupled to a cylindrical far-infrared photon cavity. The driving is introduced through a harmonic modulation of the full electron-photon interaction, therefore including both paramagnetic and diamagnetic contributions. The electron-electron Coulomb interactions are treated within density functional theory, while the electron-photon coupling is modeled using a many-body configuration interaction approach at each iteration of the density functional. By exploiting the unique properties of the cylindrical TE$_{011}$ cavity mode, we demonstrate selective enhancement of diamagnetic two-photon transitions. Our results reveal that the effectiveness of harmonic modulation of the electron-photon interaction is strongly dependent on both the driving frequency and the electron occupation number per dot. When the driving frequency approaches twice the cavity photon frequency, the system exhibits resonant behavior characterized by efficient photon pumping, occupation of higher-order photon replicas, and activation of collective radial Coulomb breathing modes. These findings establish a controllable mechanism for manipulating photon states in coupled quantum dot-cavity systems and provide insights into the interplay among harmonic modulation, photonic excitations, magnetic confinement, and many-body electron correlations in dimensionally reduced nanostructures.
\end{abstract}

\maketitle
%
%----------------------------------------------------------------------------------------
%

\section{Introduction}
Two-dimensional electron gas (2DEG) in GaAs and InAs based heterostructures has for a
long time been used to investigate electron transport, optical properties and magnetization of
structures in dimensionally reduced nano- and microstructures like quantum dots, rings
and wires besides the homogeneous 2DEG.
\cite{PhysRevLett.109.077401,Kim04:073311,PhysRevB.61.R16319,Krahne01:195303,Vasiliadou95:R8658,Gerhardts91:5192,PhysRevB.100.155301,Meinel99:819}.

In an external magnetic field the Landau energy subbands of the structures create
energy gaps or discrete levels in the conduction band that can host terahertz photon transitions
and optical effects in the far-infrared (FIR) regime \cite{HEITMANN200237,kotlyar98:3989,Dahl92:15590,Yoshie2004}.
Placing the active electron systems in different kinds of photon microcavities can
further modify their optical properties \cite{Zhang1005:2016,Maag2016,Gaur2026}.

In previous publications we have modeled static and dynamically excited 2DEG
in an array of quantum dots in external homogeneous magnetic field and embedded
in a cylindrical FIR photon-cavity with a single TE$_{011}$ mode
\cite{PhysRevB.110.205301,PhysRevB.109.235306}.
The excitation of the electron-photon system was accomplished with short
pulses modulating the electron-photon interaction. In this manner a broad
spectrum of excitation frequencies is {\lq\lq}offered{\rq\rq} to it.
The results show that the excitation mechanism modulating the
electron-photon interaction can be used to excite predominantly two-photon
diamagnetic photon transitions in the array \cite{PhysRevB.110.205301}.
Here, we investigate the continuous excitation of the system by imposing a time harmonic external
modulation driving term to the electron-photon interaction. We analyze the
dynamic population of the basis states to gain insight into the microscopic processes
in the system caused by the continuous driving. Similarly, we include a small photon loss,
via a Lindblad damping terms \cite{Lindblad76:119} to the Liouville-von Neuman equation describing the time-evolution,
to better identify one- and two-photon processes in the underlying dynamics of the system.

{We apply the so-called quantum electrodynamics density functional theory tensor-product (QED-DFT-TP)}
formalism, a density functional theory for electrons
interacting with a single cavity-photon mode, expressed in basis states constructed
by tensor products of the electron and the photon states \cite{10.1063/5.0123909}.
The formalism has been used for investigation of static and dynamic harmonically
bound electron systems without Coulomb interactions in photon cavities \cite{PhysRevB.105.115127,PhysRevB.107.235130},
and for Coulomb interacting atomic systems in a cavity \cite{10.1063/5.0123909}.
We extended this formalism to Coulomb interacting 2DEG in a periodic potential subjected to homogeneous
external magnetic field in a cylindrical cavity, both for the static and dynamic case
\cite{PhysRevB.109.235306,PhysRevB.110.205301}, and here we extend the use of the formalism to
continuously driven 2DEG systems.
{
Moreover, tensor-products of Slater determinants and
photonic Fock states were recently used by Castagnola et al.\ \cite{PhysRevResearch.6.033283}
to model time evolution of photonic and molecular observables following the excitation of the
system induced by short laser pulses. Finally, the most recent applications of this method are
presented in \cite{PhysRevA.110.043119} for the excitation of He$^+$, and in \cite{42x4-nlwy} with calculations for
nitrobenzene and light-mediated interactions between spatially separated formaldehyde molecules
embedded in an optical cavity.}

Besides our interest in the possible magnetic photon transitions, cyclotron resonance,
and their role in arrays of quantum dots in an external magnetic field we notice the relevance
of cavity fields of the cylindrical FIR-cavity to research into generating structured vector beams
with spatially nonhomogeneous transverse polarization distributions. The polarization state of a cylindrical vector beam is indeterminate on the propagation axis. Alongside vortex beams carrying non-zero orbital angular momentum, these unconventional optical beams are usually characterized by doughnut-shaped profiles in their lowest orders, and their diverse applications have led to a revolution in optical manipulation
\cite{He2022,Omatsu_2025,PhysRevA.101.023821,Zhan:09,Shen2019,Lin-10004948}.
In this context, it is interesting to keep in mind that we have earlier reported
an inhomogeneous distribution of photons in the present system \cite{PhysRevB.109.235306}.

The paper is organized as follows: In Sec.\ \ref{Model} the model is briefly described,
the results and discussion thereof are found in Sec.\ \ref{Results}, with the conclusions
drawn in Sec.\ \ref{Conclusions}. Appendix \ref{Tech-details} contains technical details
of the methodology used for the modeling.

\section{Model}
\label{Model}

The 2DEG is in a lateral square superlattice of quantum dots in a
GaAs heterostructure. The 2DEG is considered to reside in the $xy$-plane and perpendicular
to it is a homogeneous external magnetic field $\bm{B} = B\bm{e}_z$
generated by the vector potential ${\bm A} = (B/2)(-y,x)$.
In terms of the  effective mass of the electrons $m^*$,
the dielectric constant $\kappa_\mathrm{e}$, and the effective $g$-factor
$g^*$ the Hamiltonian of the 2DEG-cavity system is
\begin{equation}
	H = H_\mathrm{e} + H_\mathrm{int} + H_\gamma + H_\mathrm{ext}(t),
	\label{Hinitial}
\end{equation}
where
\begin{equation}
	H_\mathrm{e} = H_0 + H_\mathrm{Zee} + V_\mathrm{H} + V_\mathrm{per} + V_\mathrm{xc},
	\label{He}
\end{equation}
describes the 2DEG in the square array and
\begin{equation}
	H_0 = \frac{1}{2m^*}\bm{\pi}^2, \quad\mbox{with}\quad
	\bm{\pi} = \left(\bm{p}+\frac{e}{c}\bm{A} \right).
	\label{H0}
\end{equation}
The spin Zeeman term is
\begin{equation}
	H_\mathrm{Zee} = \pm g^* \mu_\textrm{B}^* B/2,
	\label{HZee}
\end{equation}
with $\mu_\textrm{B}^*$ the effective Bohr magneton.
The direct Coulomb interaction in terms of the effective local electron charge density
$\Delta n(\bm{r}) = n_\mathrm{e}(\bm{r})-n_\mathrm{b}$ is
\begin{equation}
	V_\mathrm{H}(\bm{r}) = \frac{e^2}{\kappa_\mathrm{e}}\int_{\mathbf{R}^2}d\bm{r}'\frac{\Delta n(\bm{r}')}
	{|\bm{r}-\bm{r}'|},
	\label{Vcoul}
\end{equation}
where $-en_\mathrm{e}(\bm{r})$ is the electron charge density and
$+en_\mathrm{b}$ is the homogeneous positive background charge density
representing the ionized donors securing the overall charge neutrality of the system.
The exchange correlation potential $V_\mathrm{xc}$ and the corresponding spin-density functionals are
detailed in Appendix A of Ref.\ \cite{PhysRevB.110.205301} with references to their origins.
The total energy per unit cell of the 2DEG-cavity system is likewise stated by
Eq.\ (25) in the same reference.

The square array of quantum dots is described by the periodic potential
\begin{equation}
	V_\mathrm{per}(\bm{r}) = -V_0\left[\sin \left(\frac{g_1x}{2} \right)
	\sin\left(\frac{g_2y}{2}\right) \right]^2,
	\label{Vper}
\end{equation}
where $V_0$ is the strength of the periodic potential.
$V_\mathrm{per}(\bm{r})$ is depicted in Fig.\ 1 in Ref.\
\cite{PhysRevB.108.115306}.
The superlattice is spanned by the spatial vectors
$\bm{R}=n\bm{l}_1+m\bm{l}_2$ with $n,m\in \bm{Z}$,
with the unit vectors $\bm{l}_1 = L\bm{e}_x$ and $\bm{l}_2 = L\bm{e}_y$.
The reciprocal lattice is spanned by $\bm{G} = G_1\bm{g}_1 + G_2\bm{g}_2$ with
$G_1, G_2\in \mathbf{Z}$ and the unit vectors $\bm{g}_1 = 2\pi\bm{e}_x/L$
and $\bm{g}_2 = 2\pi\bm{e}_y/L$.
The period of the superlattice is $L$, {and the primitive unit square cell defined
by the periodic potential (\ref{Vper}) is noted by $\bm{{\cal A}}$.}
The electrons of the 2DEG interact with the cavity photons according to
\begin{align}
	H_\mathrm{int} = \frac{1}{c}\int_{\mathbf{R}^2} d\bm{r}\; &
	{\bm J}({\bm r})\cdot{\bm A}_\gamma (\bm{r}) \nonumber\\
	+& \frac{e^2}{2m^*c}\int_{\mathbf{R}^2} d\bm{r}\;
	n_\mathrm{e}(\bm{r})A^2_\gamma(\bm{r}),
	\label{e-g}
\end{align}
where both the para- and the diamagnetic parts of the electron-photon interaction are
accounted for. The Hamiltonian of the cavity photons is
\begin{equation}
	H_\gamma = E_\gamma a_\gamma^\dagger a_\gamma
\end{equation}
written in terms of the  photon creation and annihilation operators
with $E_\gamma = \hbar\omega_\gamma$.
The electron-photon interaction (\ref{e-g}) becomes
\begin{align}
	H_\mathrm{int} =& g_\gamma \hbar\omega_c \left\{ lI_x + lI_y\right\} \left(a^\dagger_\gamma + a_\gamma\right)\nonumber\\
	+& g^2_\gamma \hbar\omega_c {\cal N}\left\{\left(a^\dagger_\gamma a_\gamma + \frac{1}{2}\right)
	+\frac{1}{2}\left(a^\dagger_\gamma a^\dagger_\gamma + a_\gamma a_\gamma\right)\right\}
	\label{e-gIxIyN}
\end{align}
with the integrals, $I_x$, $I_y$, and ${\cal N}$, that can be viewed as functionals
of the charge current and electron densities, respectively, defined as \cite{PhysRevB.109.235306}
\begin{align}
	l(I_x + I_y) &=  \frac{m^*}{e}\int_{\bm{{\cal A}}} d\bm{x}\;
	\frac{l}{\hbar}\left[-J_x(\bm{x})\left(\frac{y}{l}\right) + J_y(\bm{x})\left(\frac{x}{l}\right)\right]
	\nonumber\\
	{\cal N} &= \int_{\bm{{\cal A}}} d\bm{x}\; n_\mathrm{e}(\bm{x})\left(\frac{x^2+y^2}{l^2}\right).
	\label{lIxx}
\end{align}
{
This folding of the integration domains
from $\mathbf{R}^2$ to $\bm{{\cal A}}$ relies on the special form of the vector potential $\mathbf{A}_\gamma$
for the single TE011 mode of the cylindrical cavity, stated below, in the long wavelength, and the periodicity of the
system. The extra terms shown in the Appendix of Ref.\ \cite{PhysRevB.109.235306} come from summations over the unit lattice
vectors $\mathbf{R}$ expressible in Dirac-delta functions for the inverse lattice vectors $\mathbf{G}$,
see Eq.\ (A10) there.
In the analytical derivation no supercell is used and these terms represent {\lq\lq}selection rules{\rq\rq} or
conditions for the inverse lattice vectors $\mathbf{G}$ in the calculation. Of course, in the numerical
calculations there are upper limits introduced for the vectors $\mathbf{G}$, and they are determined
such that their cutoff does not influence the results in a visible way.}

All terms of the electron-photon interaction, including the ones described as the
antiresonant ones of the diamagnetic electron-photon interaction are included in the calculation
and in the present system they are
important, just as was seen for the dynamic results for the array of
quantum dots \cite{PhysRevB.110.205301} and rings \cite{PhysRevB.111.115304}
in a cylindrical photon cavity with a TE$_{011}$ mode.
The emerging dimensionless electron-photon coupling constant is
\begin{equation}
	g_\gamma = \left\{ \left( \frac{e{\cal A}_\gamma}{c} \right) \frac{l}{\hbar} \right\}.
\end{equation}
The cyclotron energy
is $E_c = \hbar\omega_c = eB/(m^*c)$, and the magnetic length is $l = [\hbar c/(eB)]^{1/2}$.
The vector potential of the TE$_{011}$ cavity mode in the long wavelength limit is
\begin{equation}
	{\bm A}_\gamma (\bm{r}) = \bm{e}_\phi {\cal A}_\gamma\left(a^\dagger_\gamma + a_\gamma  \right)
	\left(\frac{r}{l} \right),
	\label{Ag}
\end{equation}
where $\bm{e}_\phi$ is the angular unit vector in polar coordinates.
This vector potential has the same spatial form as the static vector potential
${\bm A}$ leading to the
external homogeneous magnetic field $\bm{ B} = B\bm{e}_z$ \cite{PhysRevB.109.235306}.

As Eq.\ (\ref{Ag}) depends on the spatial coordinate $r$ it promotes
magnetic transitions for the electrons. In the present system
it connects the para- and the diamagnetic electron-photon interaction
emphasizing the importance of the diamagnetic interaction in the presence of the external
magnetic field. The spatially dependent vector potential (\ref{Ag})
of the electron-photon interaction (\ref{e-g}) prevents the paramagnetic part of the
interaction to be transformed to an electron dipole interaction, and the diamagnetic part
to be approximated by a constant depending on the number of electron \cite{Gudmundsson_NJP_2026}.

{
The continuous temporal excitation of the 2DEG-cavity system {\it switched} on at $t = 0$
opens up the electron-photon system to energy and photons and  is modeled
by a harmonic time-modulation of the electron-photon coupling described by the
time-dependent Hamiltonian
\begin{align}
\label{Ht}
    H_\mathrm{ext}(t) &= F(t)
    \biggl[ g_\gamma \hbar\omega_c\left\{lI_x + lI_y \right\} \left( a^\dagger_\gamma + a_\gamma \right)\\
    + &\left. g_\gamma^2 \hbar\omega_c {\cal N} \left\{ \left( a^\dagger_\gamma a_\gamma + \frac{1}{2}\right)
    + \frac{1}{2}\left( a^\dagger_\gamma a^\dagger_\gamma + a_\gamma a_\gamma  \right) \right\} \right]\nonumber
\end{align}
with
\begin{equation}
	F(t) = \Theta(t)\left( \frac{V_t}{\hbar\omega_c}\right) \sin{\left(\omega_\mathrm{ext} t\right)},
	\label{ft}
\end{equation}
where $\Theta(t)$ is the Heaviside unit step-function and
$\omega_\mathrm{ext}$ is the frequency of the modulation of the electron-photon interaction.
In the calculations the strength of the modulation of the electron-photon interaction
is held constant, $V_t/{\hbar\omega_c} = 0.5$ together with the dimensionless electron-photon
coupling coefficient $g_\gamma = 0.06$.}

We apply the quantum electrodynamical density functional theory approach, QED-DFT-TP as presented by Malave
et al.\ \cite{10.1063/5.0123909} and adapted to our 2DEG-cavity system by calculating
the energy spectrum and the eigenstates of the static part of the total Hamiltonian
(\ref{Hinitial}) in a linear functional basis built as a tensor product (TP) of the electron and photon states
\begin{equation}
	|\bm{\alpha\theta}\sigma n\rangle = |\bm{\alpha\theta}\sigma\rangle\otimes|n\rangle.
	\label{TP}
\end{equation}
The photon states $|n\rangle$ are the eigenstates of the photon number operator, and the electron states $|\bm{\alpha\theta}\sigma \rangle$ are the single-electron eigenstates of $H_0$
proposed by Ferrari and constructed
for the periodic 2DEG in an external magnetic field at each point in the first Brillouin zone, i.e.\ $\bm{\theta} = (\theta_1,\theta_2)\in [-\pi,\pi]\times[-\pi,\pi])$
\cite{Ferrari90:4598,Silberbauer92:7355,Gudmundsson95:16744,PhysRevB.105.155302,PhysRevB.106.115308}. $\sigma\in\{\uparrow,\downarrow\}$ is the quantum number for the $z$-component of the electron spin.
All further quantum numbers of the Ferrari states \cite{Ferrari90:4598} are combined in $\bm{\alpha}$, which can be construed as a subband index of the split Landau
bands \cite{Gudmundsson95:16744,Hofstadter76:2239,Dahl90:5763,Silberbauer92:7355,Pfannkuche92:12606}.
{
In the QED-DFT-TP formalism, the electron Coulomb interaction is treated within a QED-DFT formalism (single-electron mean-field type treatment), while, by using a tensor product (TP) of the electron and the photon states of Eq.\ (\ref{TP}), the many-body system state with the electron–photon interaction is treated by exact diagonalization (configuration interaction) within each iteration step of the DFT formalism.}

{
The dynamical electron and current densities are in terms of the density operators of the
system \cite{Gudmundsson_NJP_2026}
\begin{align}
	n_\sigma (\bm{r},t) =
	\frac{1}{(2\pi)^2}\int_{-\pi}^\pi d\bm{\theta}\,
	\sum_{\bm{\alpha\beta}nn'}\phi^*_{\bm{\alpha\theta}\sigma}(\bm{r})
	\phi_{\bm{\beta\theta}\sigma}(\bm{r})\nonumber\\
	\rho^{{\bm{\theta}}}_{\bm{\alpha}n\sigma,\bm{\beta}n'\sigma}(t)
	\label{Net}
\end{align}
\begin{align}
	{J}_{i}(\bm{r},t) = \frac{-e}{m^*(2\pi)^2} \int_{-\pi}^{\pi} d\bm{\theta}\;  \sum_{\bm{\alpha\beta}\sigma nn'}
	\Re&\left\{ \phi_{\bm{\alpha\theta}\sigma}^*(\bm{r}){\pi}_i \phi_{\bm{\beta\theta}\sigma}(\bm{r}) \right\}\nonumber\\
	&\rho^{\bm{\theta}}_{\bm{\beta}n'\sigma,\bm{\alpha}n\sigma}(t)
	\label{currD}
\end{align}
respectively, for $i=x$ or $y$, and the total electron density is $n_\mathrm{e} = n_\uparrow + n_\downarrow$
in terms of the electron spin densities. 
The wavefunctions $\phi_{\bm{\alpha\theta}\sigma}(\bm{r})=\langle \bm{r} |\bm{\alpha\theta}\sigma\rangle$ correspond to the electron states $|\bm{\alpha\theta}\sigma\rangle$ used to construct the basis (\ref{TP}).
The time-dependent density matrix elements are defined as   $\rho^{{\bm{\theta}}}_{\bm{\alpha}n\sigma,\bm{\beta}n’\sigma}(t)=\langle\bm{\alpha,\theta}n\sigma|\rho(t)|\bm{\beta,\theta}n’\sigma\rangle$. The time dependence of the density matrix, or density operator, comes from its time evolution governed by the master equation (Liouville-von Neumann equation), which also includes cavity photon loss and will be presented next.}

The Liouville-von Neumann (L-vN) equation is used to describe the time evolution of the cavity-photon
dressed 2DEG
\begin{align}
\label{L-vN}
	\partial_t\rho^{\bm{\theta}} (t) = &-\frac{i}{\hbar}\left[H[\rho^{\bm{\theta}}(t)\right],\rho^{\bm{\theta}} (t)]
	\\
	&+\frac{\kappa}{2\hbar}\left\{2a_\gamma\rho^{\bm{\theta}} (t) a^\dagger_\gamma - a^\dagger_\gamma a_\gamma\rho^{\bm{\theta}} (t)
	- \rho^{\bm{\theta}} (t) a^\dagger_\gamma a_\gamma\right\}\nonumber ,
\end{align}
where a weak Lindblad-like photon dissipation has been added to the cavity controlled by the parameter $\kappa$
\cite{Gudmundsson19:10}. The dissipation is assembled by single-photon transitions assuming that the environment
of the cavity is at a low enough temperature so no photons leak into it, and will be used below
to compare the roles of single- and two-photon transitions \cite{Gudmundsson19:10}. As the spatial part
of the vector potential for the cylindrical TE$_{011}$ mode has the same type as that of the
external magnetic field, the excitation of the system with $(\ref{Ht})$ does not couple different points
in the Brillouin zone the L-vN equation (\ref{L-vN}) is solved for each point in it \cite{PhysRevB.110.205301}.

We emphasize that the time-dependent Hamiltonian in the Liouville-von Neumann equation (\ref{L-vN})
is a functional of the density operator
$\rho^{\bm{\theta}}$ through the electron charge and current densities, that have to be
updated in each iteration within each time-step of the numerical time integration of Eq.\ (\ref{L-vN}).
We employ the Crank-Nicolson scheme as it is designed for use in Hermitian systems \cite{Crank1947}.

In addition to the dynamic electron density and the current density the time-dependent
density matrices are used to evaluate the time-dependent average photon number
\begin{equation}
	N_\gamma (t) = \frac{1}{(2\pi)^2}\sum_{\sigma}\int^{\pi}_{-\pi} d\bm{\theta}\; \mathrm{Tr} \left\{ \rho^{\bm{\theta}}_{\sigma}(t) a^\dagger_\gamma a_\gamma\right\},
	\label{Ngt}
\end{equation}
and the dynamical orbital magnetization \cite{Desbois98:727,Gudmundsson00:4835}
\begin{equation}
	Q_J(t) = \frac{1}{2c{\cal A}}\int_{\cal A} d\bm{r} \left( {\bf r}\times
	\langle {\bf J}({\bf r},t) \rangle \right) \cdot{\bm{e}_z},
	\label{Mo}
\end{equation}
with ${\cal A} = L^2$. The number of electrons in each dot or unit cell is noted by $N_\mathrm{e}$.
We calculate the mean values, $Q_0 = \langle x^2+y^2 \rangle$, $Q_1 = \langle x\rangle$ and $\langle y\rangle$,
and $Q_2 = \langle xy\rangle$. These mean values are calculated
directly from the electron time-dependent density and supply information about,
breathing or monopole density oscillations, quadrupole modes, and dipole modes excited in the
system, respectively.

We acknowledge that the cavity with a single TE$_{011}$ photon-mode in the long wavelength approximation
interacting with a 2DEG represents a very special case as the description above has made
clear. We lay no claim to having established a general approach to QED-DFT-TP for a periodic 2DEG in a
homogeneous magnetic field. What strengthens our approach is the possible {\lq\lq}folding{\rq\rq} of the
integrals $I_x$, $I_y$, and ${\cal N}$ for the electron-photon interaction (\ref{e-gIxIyN}) from $\bm{R}^2$ to the unit cell $\bm{{\cal A}}$. Furthermore, it is encouraging to see findings about the commensurability conditions
in magnetic field in cavity systems \cite{PhysRevLett.123.047202}, and a study of the interaction
of photons with classical currents \cite{Dick_2024}.  The construction of the basis states (\ref{TP})
and thus the photon replicas is carried out in analogy with single-electron systems. A different
approach is presented in Ref.\ \cite{nielsen2019dressedorbitalapproachcavityquantum,10.1021/acs.jctc.0c00469}.
We stress that in our approach the choice of the basis states (\ref{TP}) representing Landau-subband polaritons,
or cavity-photon dressed electrons states, means that we can not within the present scheme describe what happens
when the electron photon-coupling vanishes, $g_\gamma \rightarrow 0$, and the electron states make the limit
of zero magnetic field in a similar way inaccessible. 
%

%------------------------
\section{Results}
\label{Results}

In the numerical calculations, we assume GaAs parameters, the effective electron mass
$m^* = 0.067m_e$, the dielectric constant $\kappa_\mathrm{e} = 12.4$, and the effective
$g$-factor $g^* = -0.44$.
The magnetic flux through a square unit cell of area $L\times L$ is according to the commensurability condition
$B{\cal A}= BL^2 =pq\Phi_0$ with $p$ and $q$ integers
\cite{Ferrari90:4598,Silberbauer92:7355,Gudmundsson95:16744},
and the magnetic flux quantum $\Phi_0 = hc/e$. We select $pq = 1$ corresponding to the external
magnetic field $B = 0.4138$ T and the cyclotron energy $E_c=\hbar\omega_c = 0.7145$ meV.
In the static part of the calculations we assume the temperature $T = 1$ K.
The rather low value for the magnetic field, $B = 0.4138$ T, yields an effective magnetic length
$l=39.89$ nm that has to be compared to the lattice length $L = 100$ nm, and the
effective Bohr radius $a^*_B=9.79$ nm corresponding to the Rydberg energy $R^*_y=5.92$ meV.
The strength of the periodic potential is set $V_0 = 16$ meV.

\subsection{H-like array ($N_\mathrm{e} = 1$)}
\label{Ne1-Results}

\begin{figure*}[htb]
	\centerline{\includegraphics[width=0.42\textwidth]{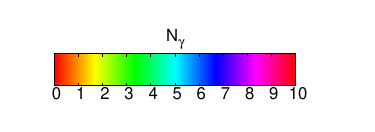}}
	\vspace*{-2.0cm}
	\begin{center}
    \includegraphics[width=0.38\textwidth,bb = 0 50 190 340]{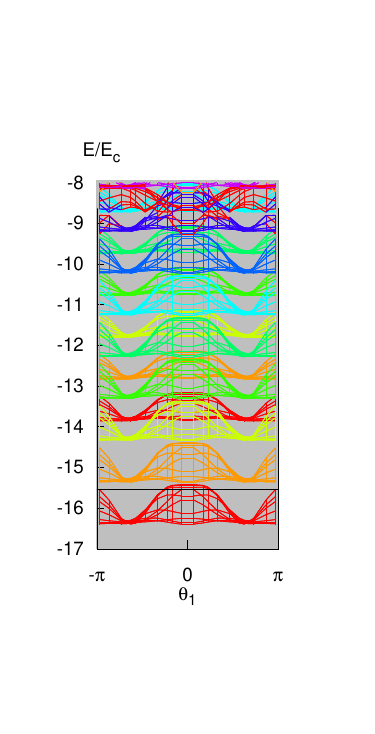}
    \includegraphics[width=0.38\textwidth,bb = 0 50 190 340]{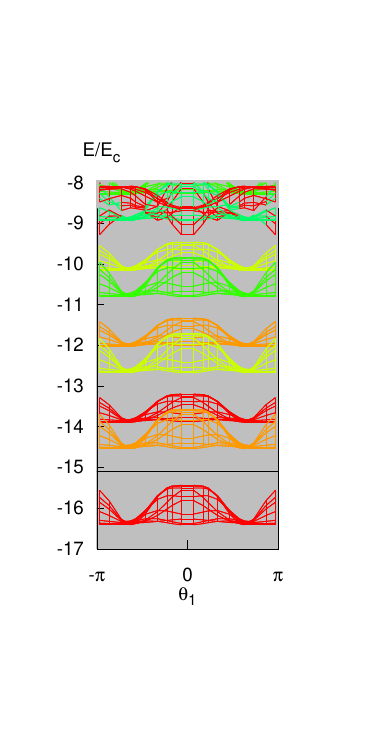}
    \end{center}
    \caption{Energy bandstructure projected on the $\theta_1$ axis in reciprocal space
             for (left) $E_\gamma = 0.7$ meV, (right) $E_\gamma = 1.3$ meV. The photon content of the
             energy bands is color coded according to the scale at the top.
             The black horizontal line indicates the chemical potential.
             The vertical colored lines correspond to the axis $\theta_2$ which is perpendicular to the
             axis $\theta_1$.
             $N_\mathrm{e} = 1$, $g_\gamma = 0.06$, $pq = 1$ corresponding to $E_c = \hbar\omega_c = 0.7145$ meV,
             $V_0 = 16.0$ meV, $L = 100$ nm, and $T = 1$ K.}
\label{1e-spectra}
\end{figure*}

Figure~\ref{1e-spectra} shows the energy spectrum for one electron in each unit cell,
$N_\mathrm{e}=1$, and for photon energies $E_\gamma = 0.7$\,meV (left) and
$E_\gamma = 1.3$\,meV (right). The subbands have a considerable width compared to their ``separation''.
This width reflects lattice effects facilitated by the shallow potential of the quantum dots and
the long effective magnetic length. To make the bandstructure clear, the surface of each subbands is made transparent,
except for the lines tracing them. Only the lowest part of the actual energy subband structure in
the calculations is shown in Fig.~\ref{1e-spectra} for clarity.

As the basis $\{|\bm{\alpha\theta}n\sigma\rangle\}$ in which the electron-photon Hamiltonian is diagonalized is a tensor product of electron and photon states, the resulting cavity-photon dressed electron Landau subbands act like photon replica bands. The photon number is an integer for each state in the initial electron-photon basis $\{|\bm{\alpha\theta}n\sigma\rangle\}$ and its energy is shifted by $nE_\gamma$ compared to the energy
of the original pure electron state, but in the resulting states of the ``interacting'' subbands $|\bm{\alpha\theta}\sigma)$ it is not necessarily so. The interactions mix up states with different photon numbers.
Therefore, the electron Landau subbands have a mean value of photon content at each point of the Brillouin zone, and this mean photon content is indicated by the color scale shown at the top of Fig.~\ref{1e-spectra}. Note that the
mean photon number $N_\gamma$ defined in Eq.\ (\ref{Ngt}) is averaged over the basis states
$|\bm{\alpha\theta} \sigma n\rangle$ (\ref{TP}) and the mean photon number at each point in the Brillouin zone $N^{\bm{\theta}}_\gamma$ corresponding to the integrand in Eq.\ (\ref{Ngt}) is color coded into the
energy subbands in Fig.\ \ref{1e-spectra}.

For the case of one electron in each unit cell, the exchange Coulomb force is strong enough to enhance the spin splitting of the subbands, so only one spin subband is below the chemical potential, the black horizontal line in Fig.~\ref{1e-spectra}. The other spin band corresponding to the occupied band is the first red subband above the chemical potential, indicating zero or very low photon content. Without exchange enhancement, the Zeeman energy splitting of the two spin projections in GaAs for $B = 0.4138$\,T is only $0.0105$\,meV, a value that would not be easily discernible on the energy scale of Fig.~\ref{1e-spectra}. The exchange force is therefore clearly strong and cannot be neglected. It is basically an inter-dot exchange force, showing that, for the selected parameters, the interdot Coulomb forces are important \cite{PhysRevB.56.9707}.
Due to the rather low photon energy, the next band above the chemical potential is the first photon replica of the occupied band below it. The photon replicas close to the top of the figure show resonances, or interaction effects, caused by the periodic lattice potential and the electron-photon interaction given by Eq.\ (\ref{e-g}).

The energy subbands in Fig.\ \ref{1e-spectra} are two dimensional in the $\bm{\theta}=(\theta_1,\theta_2)$
reciprocal space, but the figure shows their projection onto the $\theta_1$ variable.
For the parameters selected, the photon replicas close to the top of the figure show resonances,
or interaction effect, caused by the periodic lattice potential and the electron-photon
interaction (\ref{e-g}).

Placing the modulated 2DEG system in a photon cavity introduces photon replicas into the
energy bandstructure of the system. These are subsequently modified by the lattice effects
of the superlattice and the Coulomb forces between the electrons.
FIR-photon transitions between the replicas spurred by the continuous modulation of the
electron-photon interaction~(\ref{Ht}) serve two goals. First, they can be used to drive
the system into an excited state with a high photon number, and second, the analysis of
the collective oscillations excited in the system gives information about microscopic processes in it.

\begin{figure}[htb]
    \includegraphics[width=0.48\textwidth]{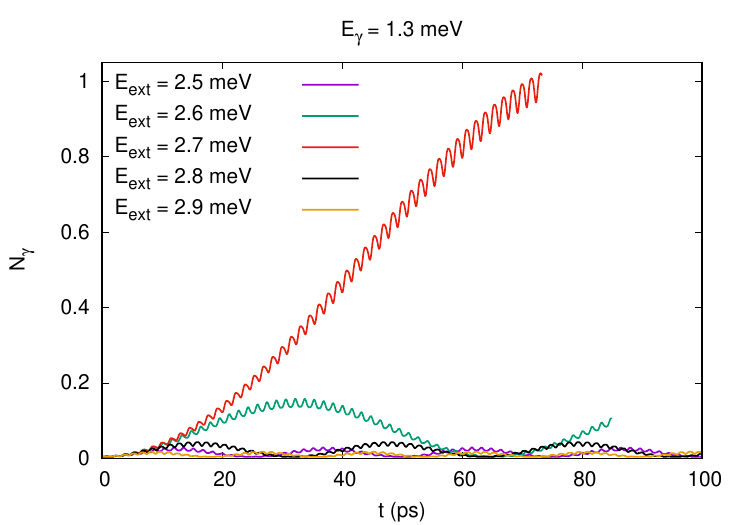}
    \includegraphics[width=0.48\textwidth]{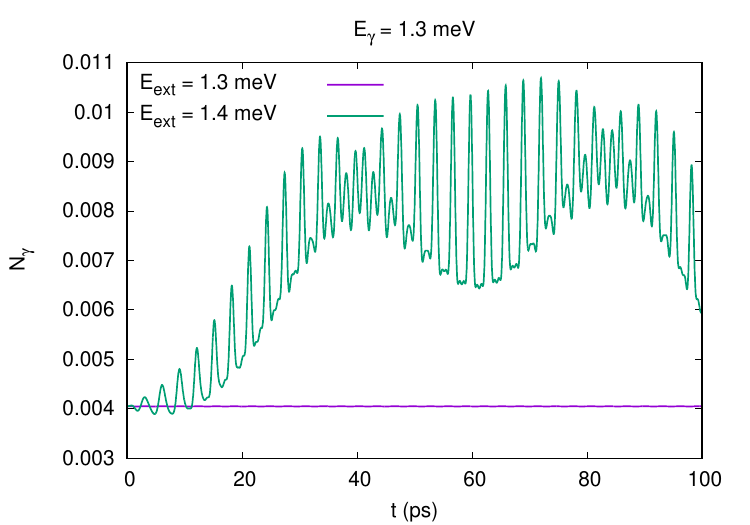}
    \includegraphics[width=0.48\textwidth]{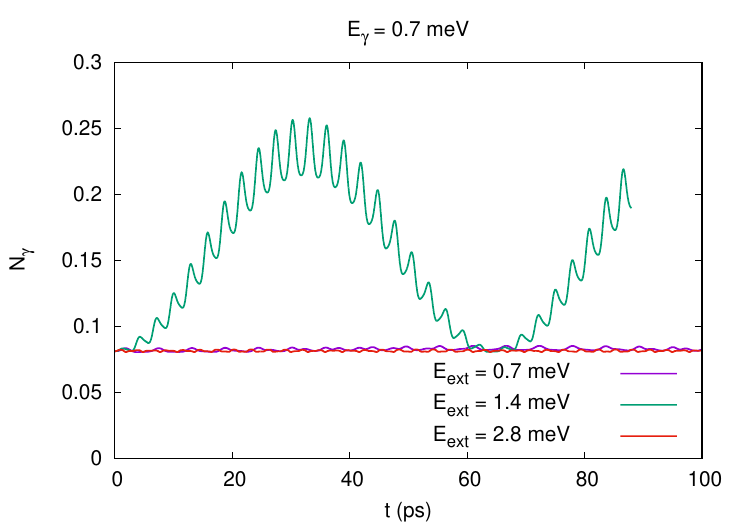}
    \caption{The mean photon number $N_\gamma$ as function of time for various
             values of the photon energy $E_\gamma = \hbar\omega_\gamma$ and
             modulation frequency $E_\mathrm{ext} = \hbar\omega_\mathrm{ext}$ of
             the electron-photon interaction. $N_\mathrm{e} = 1$, $g_\gamma = 0.06$, $pq = 1$,
             $V_0 = 16.0$ meV, $L = 100$ nm, $V_t/\hbar\omega_c = 0.5$, and $T = 1$ K.}
    \label{1e-Ng}
\end{figure}

We use our experience with exciting the system with a short time modulation pulse of the electron-photon
interaction to try the cases presented in Fig.\ \ref{1e-Ng} \cite{PhysRevB.110.205301}.
The aim is both to find a photon energy $E_\gamma$ and a modulating driving frequency (energy)
$\hbar\omega_\mathrm{ext}$ that can be used to increase the photon content of the system.
In doing so we are aware that we can not select exact resonance cases as they would
require a large Hilbert space of basis elements to correctly describe the situation, and at the
same time we needed to explore the sensitivity of the system to the parameter choices without
a large restrain on the available CPU-time. We select
$E_\gamma = 1.3$ or 0.7 meV bearing in mind that $\hbar\omega_c = 0.7145$ meV, and thus
$E_\gamma/\hbar\omega_c \approx 1.82$ or 0.98, respectively. These ratios will be more important
below when the Fourier power spectra for the results are analyzed.

The top subfigure in Fig.\ \ref{1e-Ng} displays the mean photon number per unit cell
of the 2DEG cavity system as a function of time for $E_\gamma = 1.3$ meV and several values of
$\hbar\omega_\mathrm{ext}$. Clearly, the most effective accumulation of photons happens for
$\hbar\omega_\mathrm{ext} = 2.7$ meV, indicating that close to the corresponding frequency
driving the modulation of the
electron-photon interaction the system could be for some time continuously pumped to
higher photon levels. Off-tuning of this driving frequency leads to continued oscillation of
$N_\gamma$. This behavior is in accordance with what can be found for a undamped classical driven
oscillator with or without internal structure. The system is open in the sense that the driving
term can supply or extract energy from it as can be seen by the mean photon number or the
time dependent total energy (not shown here).

Similar behavior can be found for the lower photon energy $E_\gamma = 0.7$ meV seen in the lowest
subfigure of Fig.\ \ref{1e-Ng}, where clearly the driving energy $\hbar\omega_\mathrm{ext} = 1.4$ meV
is close to a resonance for increasing $N_\gamma$, but for $\hbar\omega_\mathrm{ext} = E_\gamma$
or $\hbar\omega_\mathrm{ext} = 3E_\gamma$ the amplitude of the photon oscillations is very small.
The sensitivity of the photon pumping to the exact value of $\hbar\omega_\mathrm{ext}$ is well seen
in the center subfigure of Fig.\ \ref{1e-Ng}. Note that the subfigures have different $N_\gamma$ scales.
Here, we have thus seen, like will become evident for other values of $N_\mathrm{e}$ below, that the
photon content of the quantum dots in the current system is best changed when the driving energy
is approximately equal to twice the photon frequency of the cylindrical cavity.

\begin{figure*}[htb]
    \includegraphics[width=0.48\textwidth]{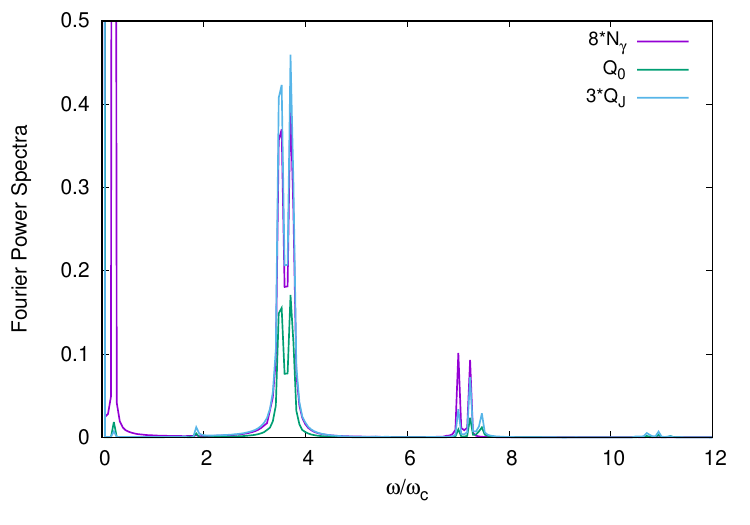}
    \includegraphics[width=0.48\textwidth]{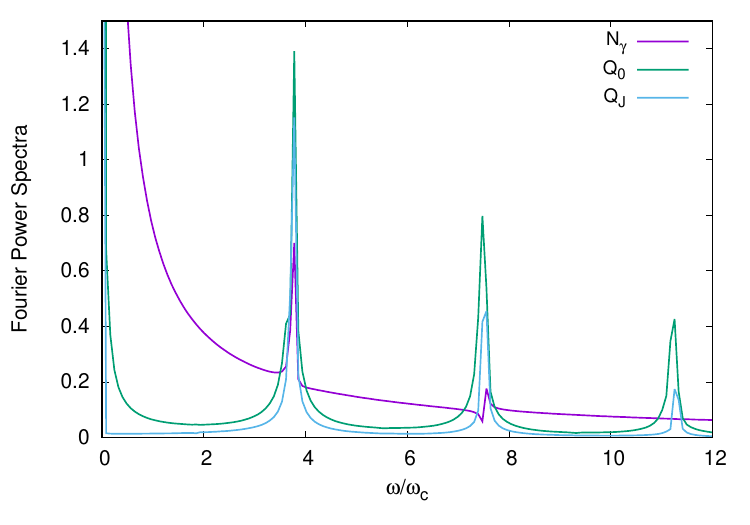}
    \includegraphics[width=0.48\textwidth]{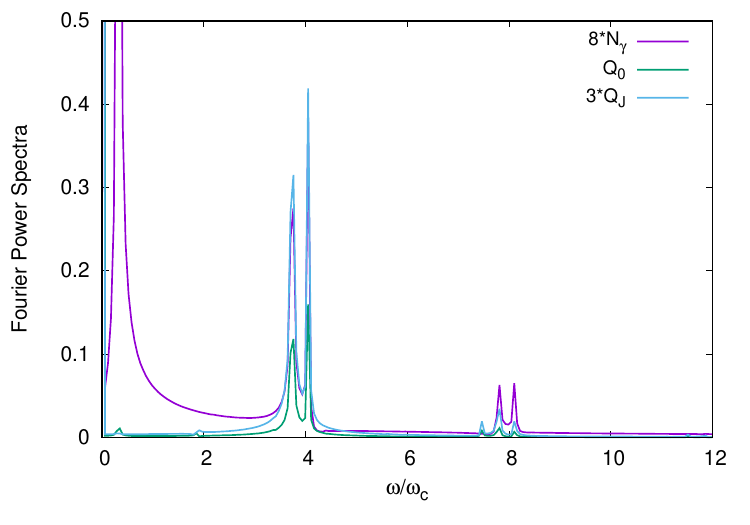}
    \includegraphics[width=0.48\textwidth]{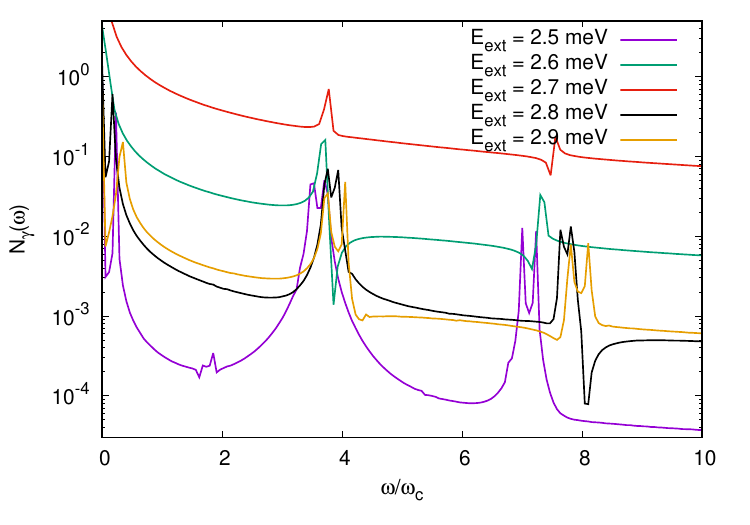}
    \caption{The Fourier power spectra $N_\gamma (\omega)$, $Q_0(\omega)$, and
             $Q_J(\omega)$ for $E_\mathrm{ext} = \hbar\omega_\mathrm{ext} =
             2.5$ meV (top left), 2.7 meV (top right), and 2.9 meV (bottom left), and
             $N_\gamma (\omega)$ compared for several values of $E_\mathrm{ext}$ on a
             logarithmic scale. $N_\mathrm{e} = 1$, $E_\gamma = 1.3$ meV, $g_\gamma = 0.06$, $pq = 1$,
             corresponding to $E_c = \hbar\omega_c = 0.7145$ meV, $V_t/\hbar\omega_c = 0.5$, and $T = 1$ K.}
    \label{1e-Eg13-t-spectra}
\end{figure*}

A clearer understanding of the situation can be obtained from analyzing the Fourier power spectra
for $N_\gamma$, $Q_0$, and $Q_J$ presented in Fig.\ \ref{1e-Eg13-t-spectra}.
Not surprisingly the simplest and the strongest peak structure is seen for the resonance
condition, when $\hbar\omega_\mathrm{ext} = 2.7$ meV in the upper right subfigure of
Fig.\ \ref{1e-Eg13-t-spectra}. The main (and the lowest) peak of $N_\gamma(\omega)$ occurs
just below $\omega/\omega_c = 4$, and above we saw that $E_\gamma/\hbar\omega_c \approx 1.82$.
This peak is thus due to two-photon diamagnetic processes as was also seen in when the system
was excited with short modulation pulses \cite{PhysRevB.110.205301}. A weak second harmonic is seen
for $N_\gamma$.  Furthermore, we note clearly visible
fundamental and higher harmonics of the peaks of the dynamic orbital magnetization $Q_J(\omega)$
and the radial breathing mode $Q_0(\omega)$.

This can be explained by the special properties of the TE$_{011}$ mode in cylindrical cavities.
The radial spatial dependence of the vector field $\bm{A}_\gamma$ makes the diamagnetic electron-photon
interaction to be sensitive to the radial shape of the electron charge density, but not only on the
number of electrons in each dot $N_\mathrm{e}$. The Lorentz force of the external magnetic field
(represented by $\bm{A}$ having the same spatial dependence as $\bm{A}_\gamma$ in
the long wavelength approximation) couples the radial and the rotational movements of
the electron charge and thus also the para- and the
diamagnetic parts of the electron-photon interaction \cite{Gudmundsson_NJP_2026}.

In the two left subfigures of Fig.\ \ref{1e-Eg13-t-spectra} we see a weaker split peak structure
when $\hbar\omega_\mathrm{ext}$ is detuned to 2.5 (top) and 2.9 (bottom). In addition we note two
low frequency peaks in $N_\gamma(\omega)$ representing the envelop of the time oscillations
$N_\gamma(t)$. This last fact explains the broad structure seen in $N_\gamma(\omega)$ at the low
frequency end when $\hbar\omega_\mathrm{ext} = 2.7$ meV, the system is not exactly at a resonance
condition between $\hbar\omega_{\text{ext}}$ and $E_\gamma$.
(top-right subfigure of Fig.\ \ref{1e-Eg13-t-spectra}). Together with the peak splitting at
$\hbar\omega_\mathrm{ext} = 2.5$ or 2.9 meV we notice a slight frequency shift, that is especially
clear for the second harmonics when compared on a logarithmic $N_\gamma$ scale in the bottom
right panel of Fig.\ \ref{1e-Eg13-t-spectra}.

The main peak just below $\omega/\omega_c \approx 4$ represents 2-photon diamagnetic transitions
develops a side peak caused by the excitation energy $E_\mathrm{ext}$ when it is detuned from the
twofold photon energy $2E_\gamma$. The second harmonic of this structure located below
$\omega/\omega_c \approx 8$ becomes more complex as the photon excitation are then into photon replica
subbands that are affected by interactions with the lattice structure of the array as can be seen from the
tops of the energy bandstructure presented in Fig.\ \ref{1e-spectra}.

\begin{figure}[htb]
	\includegraphics[width=0.48\textwidth]{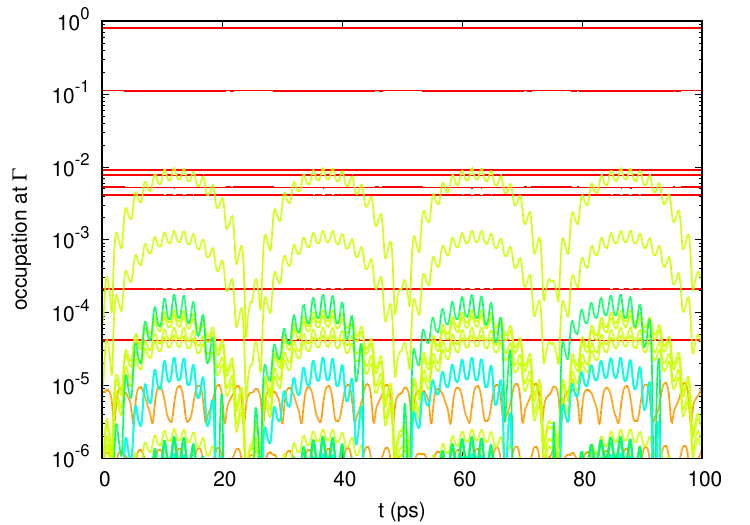}
	\includegraphics[width=0.48\textwidth]{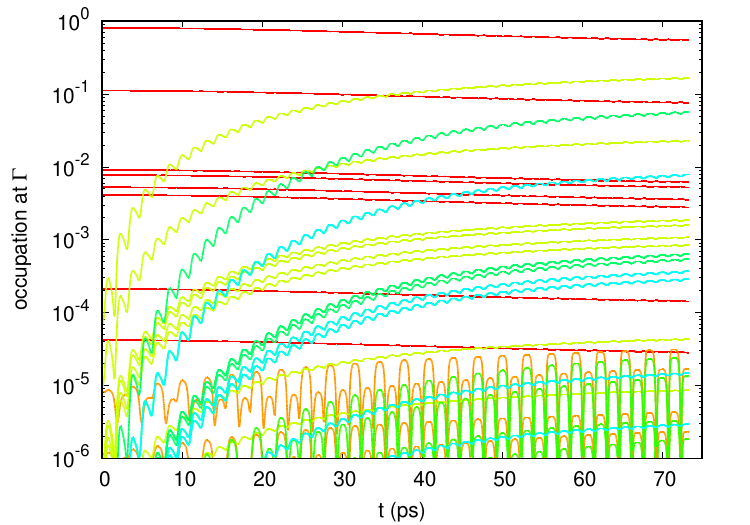}
	\caption{The occupation of the states $|\bm{\alpha\theta}n\sigma\rangle$ versus
	         time for (top) $E_\mathrm{ext} = 2.5$ meV, and (bottom) $E_\mathrm{ext} = 2.7$ meV.
	         The photon content of the states is color coded according
	         to the scale used in Fig.\ \ref{1e-spectra}.
	         $N_\mathrm{e} = 1$, $E_\gamma = 1.3$ meV, $g_\gamma = 0.06$, $pq = 1$,
	         corresponding to $E_c = \hbar\omega_c = 0.7145$ meV, $V_t/\hbar\omega_c = 0.5$, and $T = 1$ K.}
	\label{1e-occ-Comp}
\end{figure}

In order to gain further insight into the microscopic processes at work in the system
during the excitation we analyze the diagonal elements of the density matrix of system $\rho^{\bm{\theta}}(t)$
close to the $\Gamma$-point in reciprocal space in Fig.\ \ref{1e-occ-Comp}.

The properties of the static system and the time-evolution of the electron-photon system are calculated
using the {\lq\lq}noninteracting{\rq\rq} states $|\bm{\alpha\theta}n\sigma\rangle$ as a basis.
Fig.\ \ref{1e-spectra}, on the other hand, shows the energy bandstructure and photon content of the
{\lq\lq}interacting{\rq\rq} states $|\bm{\alpha\theta}\sigma)$. In Fig.\ \ref{1e-occ-Comp}
the {\lq\lq}occupation{\rq\rq} of the basis states $|\bm{\alpha\theta}n\sigma\rangle$ is shown
close to the $\Gamma$-point of the interacting system. This could have been transformed
into the interacting states, but this point of view gives a clearer picture of what is happening
with respect to the photon number. The information is the occupation of the $|\bm{\alpha\theta}n\sigma\rangle$
states, but to be exact the transformation to the interacting states would mix up the occupation and the contribution of the noninteracting states to the interacting states.

In the upper subfigure of Fig.\ \ref{1e-occ-Comp} for $\hbar\omega_\mathrm{ext} = 2.5$ meV we can identify,
by using the color code for the photon content of the states in Fig.\ \ref{1e-spectra}, that mainly
states with no photon content are occupied, or contribute to the occupied states of the system.
With a much lower probability we can identify 2-photon states with two superimposed oscillations (yellow).
1-photon states (orange) can only be identified with much lower occupation and a single frequency.
This supports the view mostly diamagnetic 2-photon processes are activated by the excitation (\ref{Ht}).

In the lower subfigure of Fig.\ \ref{1e-occ-Comp} we see what happens close to the photon pumping resonance
as $\hbar\omega_\mathrm{ext} = 2.7$ meV. Clearly the occupation of states with no photon content is reduced
as time passes and instead we see states with 2 and 4 photons becoming occupied. Again the 1-photon
states only slightly become occupied at a very low level. Not clearly seen in Fig.\ \ref{1e-occ-Comp}
are small oscillations of the highest occupied states with no photon content. These oscillations represent
radial or breathing charge oscillations (monopole) initially excited by the purely photonic excitation of the
system, but then maintained by the confining potential of each dot and interdot direct and exchange Coulomb
forces. A better opportunity to analyze these charge oscillations will be presented below as
the number of electrons in each dot $N_\mathrm{e}$ is increased.
We stress here that the photonic excitation mechanism built on the TE$_{011}$ cylindrical modes (\ref{Ht}) does
not activate any center of mass charge or electrical quadrupole modes. This is monitored through all
the calculations by evaluating $Q_1$, $Q_2$, and the induced charge density of the dynamic system.

\begin{figure*}[htb]
	\includegraphics[width=0.48\textwidth]{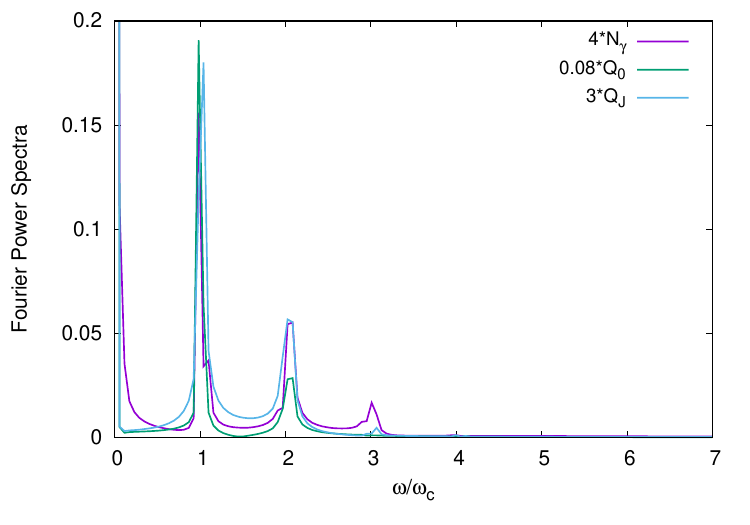}
	\includegraphics[width=0.48\textwidth]{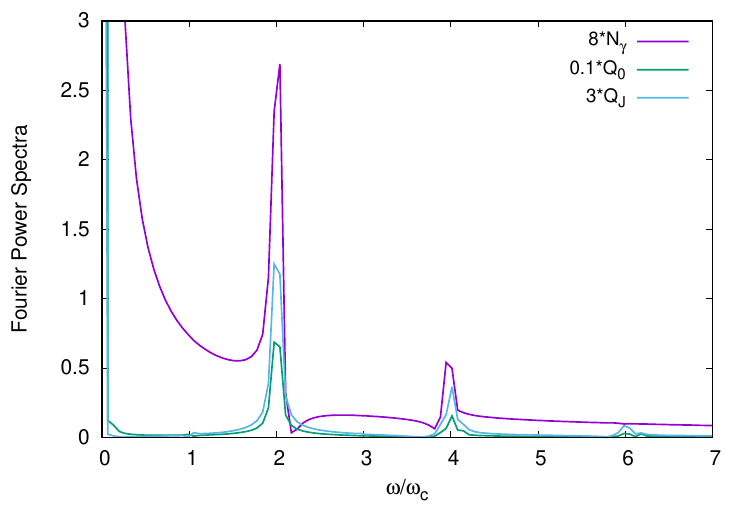}
	\includegraphics[width=0.48\textwidth]{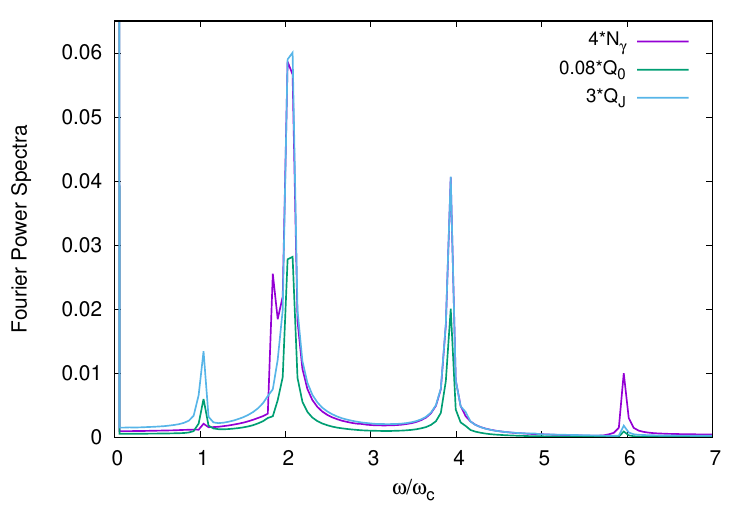}
	\includegraphics[width=0.48\textwidth]{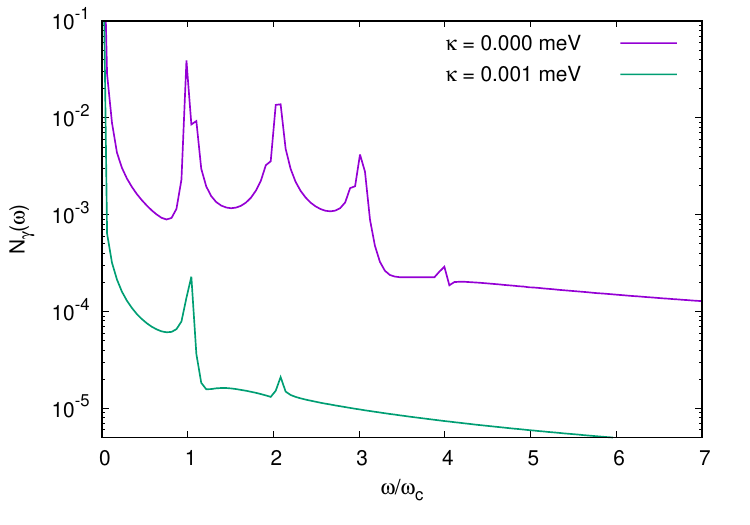}
	\caption{The Fourier power spectra $N_\gamma (\omega)$, $Q_0(\omega)$, and
		$Q_J(\omega)$ for $E_\mathrm{ext} = \hbar\omega_\mathrm{ext} =
		0.7$ meV (top left), 1.4 meV (top right), and 2.8 meV (bottom left), and
		$N_\gamma (\omega)$ compared for two values of the dissipation $\kappa$ on a
		logarithmic scale (bottom right) for $E_\gamma = 0.7$ meV and
		$E_\mathrm{ext} = 0.7$ meV corresponding to the upper left panel.
		$N_\mathrm{e} = 1$, $E_\gamma = 0.7$ meV, $g_\gamma = 0.06$, $pq = 1$,
		corresponding to $E_c = \hbar\omega_c = 0.7145$ meV, $V_t/\hbar\omega_c = 0.5$, and $T = 1$ K.}
	\label{1e-Eg07-t-spectra}
\end{figure*}

Before analyzing the behavior of the system for a higher number of electrons in a quantum
dot it is important to investigate the system with a parameter choice corresponding to
the top panel of Fig.\ \ref{1e-Ng}, i.e.\ for $N_\mathrm{e} = 1$, and $E_\gamma = 0.7$ meV.
Fig.\ \ref{1e-Eg07-t-spectra} shows the corresponding Fourier power spectra for
$E_\mathrm{ext} = 0.7$, 1.4, and 2.8 meV, $E_\mathrm{ext}/\hbar\omega_c \approx 0.98$, 1.96, and
3.9, respectively.
For, the case of $E_\mathrm{ext} = 1.4$ meV, when the system is close to a resonant driving,
we see peak structure in the top right panel similar to the one observed for the resonant
case with $E_\gamma = 1.3$ meV in the top right panel of Fig.\ \ref{1e-Eg13-t-spectra},
except now the peaks are located at $\omega/\omega_c \approx 2$, and 4, corresponding to
2-photon diamagnetic processes and the second harmonic thereof.

For the weak response at $E_\mathrm{ext} = 0.7$ meV (upper left panel of Fig.\ \ref{1e-Eg07-t-spectra})
the main peak corresponds to 1-photon paramagnetic processes located at $\omega/\omega_c \approx 1$.
The second peak, to largest extent, is caused by 2-photon diamagnetic transitions.
For $E_\gamma = 0.7$ meV and $E_\mathrm{ext} = 2.8$ meV (bottom left) the main peak (the second peak)
is caused by 2-photon diamagnetic transitions. In addition, a weak paramagnetic 1-photon peak is
also visible. The high value of $E_\mathrm{ext}$ leads to a relatively strong second harmonic peak
for the 2-photon diamagnetic transitions just below $\omega/\omega_c\approx 4$ even though the
total response of the system is weak.

The right bottom panel of Fig.\ \ref{1e-Eg07-t-spectra} displays $N_\gamma(\omega)$ with and without
a slight dissipation. Without dissipation $N_\gamma(t)$ slightly increases with time, but mainly shows
oscillations, but with the dissipation $N_\gamma(t)$ decreases by a very tiny amount with oscillations imposed. With dissipation the slight occupation increase of 1-photon states is missing and the oscillations of the
1- and 2-photon states almost vanishes (not shown here), the small dissipation effectively counteracts
the photon pumping in this weak excitation case.

For all the above cases $g_\gamma = 0.06$, but the addition of photon replicas into the
energy band structure with different photon energy $E_\gamma$ varies the effective electron-photon
interaction strength. Not surprisingly, strong effective electron-photon coupling promotes
diamagnetic transitions in the system and their role can be largely enhanced by selecting
$E_\mathrm{ext}$ to be double the photon energy. Weak effective coupling and $E_\mathrm{ext}$ detuned to
be below the double photon energy leads to the appearance of peaks reflecting 1-photon paramagnetic
transitions in the system.

\subsection{He-like array ($N_\mathrm{e} = 2$)}
\label{Ne2-Results}

\begin{figure*}[htb]
	\centerline{\includegraphics[width=0.42\textwidth]{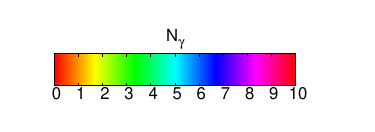}}
	\vspace*{-2.0cm}
	\begin{center}
		\includegraphics[width=0.38\textwidth,bb = 0 50 190 340]{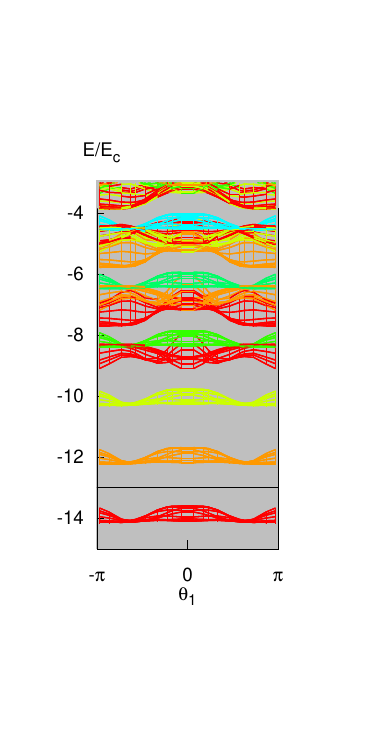}
		\includegraphics[width=0.38\textwidth,bb = 0 50 190 340]{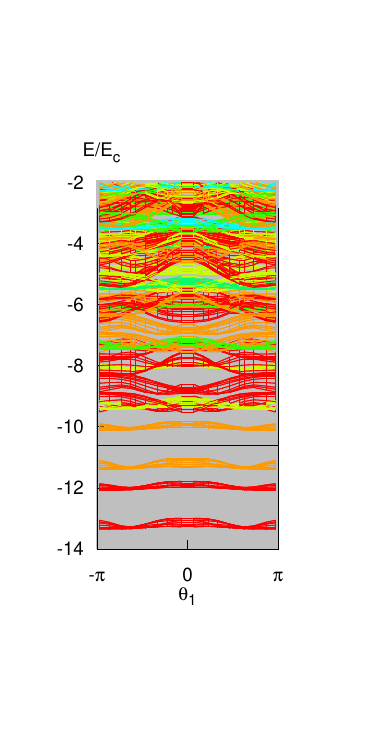}
	\end{center}
	\caption{Energy bandstructure projected on the $\theta_1$ axis in reciprocal space
		for (left) $N_\mathrm{e}=2$, $E_\gamma = 1.3$ meV, (right) $N_\mathrm{e}=3$,
		$E_\gamma = 1.3$ meV. The photon content of the
		energy bands is color coded according to the scale at the top.
		The black horizontal line indicates the chemical potential.
		The vertical colored lines correspond to the axis $\theta_2$ which is perpendicular to the
	    axis $\theta_1$.
		$g_\gamma = 0.06$, $pq = 1$ corresponding to $E_c = \hbar\omega_c = 0.7145$ meV,
		$V_0 = 16.0$ meV, $L = 100$ nm, and $T = 1$ K.}
	\label{23e-spectra}
\end{figure*}

In Fig.\ \ref{23e-spectra} the energy bandstructure is displayed for
$N_\mathrm{e}=2$ (left) and $3$ (right) for photon energy $E_\gamma = 1.3$ meV, and
$pq=1$. For $N_\mathrm{e}=2$ the electrons are in a spin-singlet state
in each dot and the Zeeman energy (or the spin splitting) is small enough
not to be easily resolved in the figure. The two occupied spin subbands below the
chemical potential seem to overlap completely in energy.

For $N_\mathrm{e}=3$ many-electron state is close to a spin triplet configuration
one spin unpaired, i.e.\ close to a Li-like configuration, compared to the He-like
situation seen for $N_\mathrm{e}=2$. This leads to strong exchange forces that are
reflected in the enhanced spin splitting visible between the two lowest subbands for
$N_\mathrm{e}=3$, which are of opposite spin. For the photon energy $E_\gamma = 1.3$ meV
the lowest photon replica of the lowest energy subband becomes the highest occupied
subband, just below the chemical potential. The occupation of a photon replica just
below the chemical potential and a small energy gap make the Li-like dot system
a good candidate for strong effective electron-photon coupling.

The width of the occupied subbands in Figures \ref{1e-spectra} and \ref{23e-spectra}
shows how the screening effects of the 2DEG increase with the electron number $N_\mathrm{e}$.
Note that the energy scales in the two figures cover the same energy span though they start
with different values.

\begin{figure}[htb]
	\includegraphics[width=0.48\textwidth]{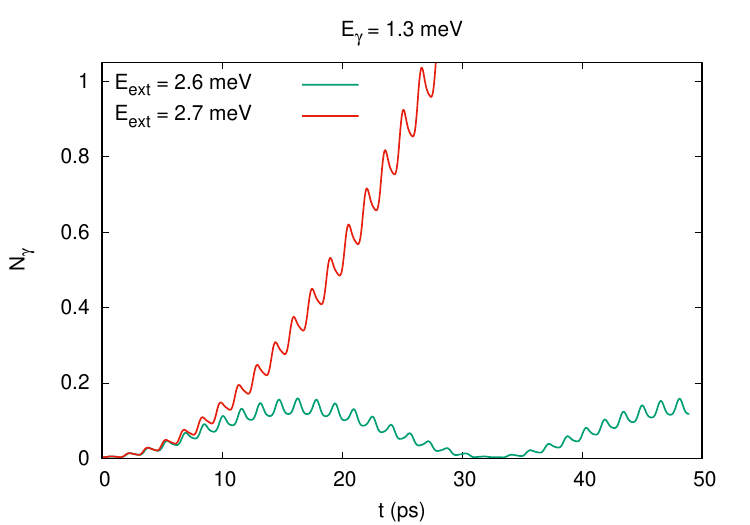}
	\caption{The mean photon number as function of time for two values of
		$E_\mathrm{ext}$ and $N_\mathrm{e} = 2$. $E_\gamma = 1.3$ meV, $g_\gamma = 0.06$, $pq = 1$,
		corresponding to $E_c = \hbar\omega_c = 0.7145$ meV, $V_t/\hbar\omega_c = 0.5$, and $T = 1$ K.}
	\label{2e-Ng-13}
\end{figure}

Figure \ref{2e-Ng-13} shows $N_\gamma(t)$ for $N_\mathrm{e} = 2$, $pq = 1$,
$E_\gamma = 1.3$ meV, and $E_\mathrm{ext} = 2.6$ and $2.7$ meV. Clearly, again we notice that
the photon number is effectively increased for $E_\mathrm{ext} = 2.7$ meV.
Generally, the number of iterations needed to obtain convergence in the time integration of
the coupled set of L-vN equations for the array of 2-electron quantum dots is higher than
for 1-electron dots, specially close to an excitation resonance. This comes from the
increased number of internal processes including Coulomb forces and leads to shorter
time series for the results.

\begin{figure*}[htb]
	\includegraphics[width=0.48\textwidth]{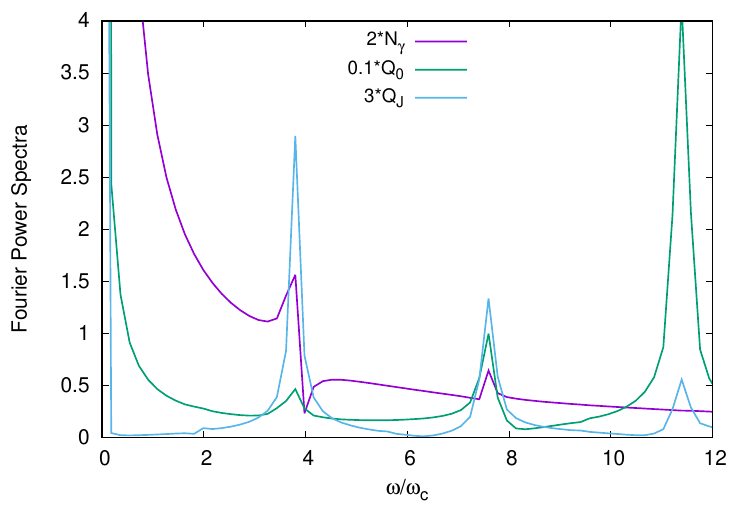}
	\includegraphics[width=0.48\textwidth]{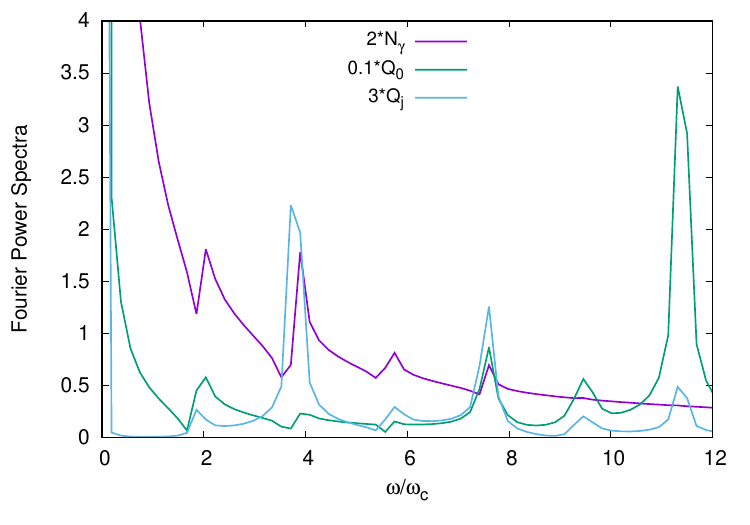}
	\includegraphics[width=0.48\textwidth]{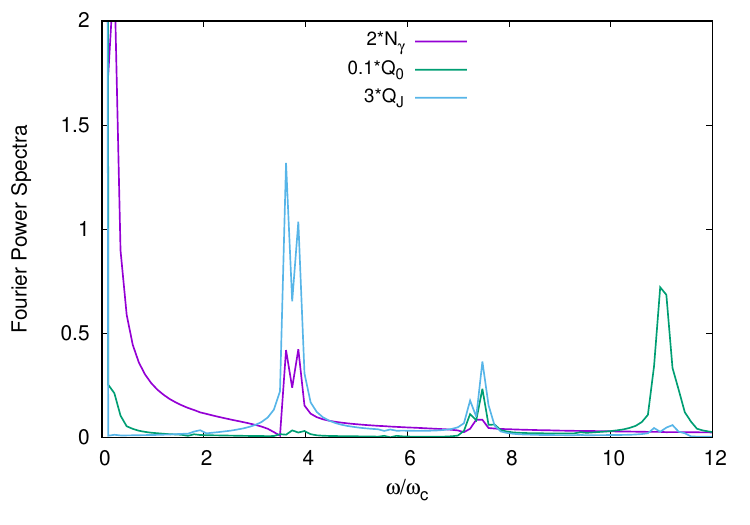}
	\includegraphics[width=0.48\textwidth]{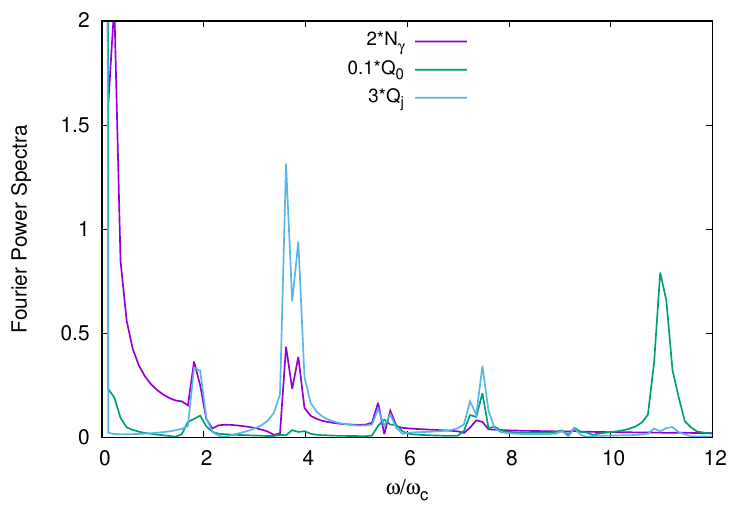}
	\caption{The Fourier power spectra $N_\gamma (\omega)$, $Q_0(\omega)$, and
		$Q_J(\omega)$ for $E_\mathrm{ext} = \hbar\omega_\mathrm{ext} =
		2.7$ meV and $\kappa = 0.0$ meV (top left), $E_\mathrm{ext} = 2.7$ meV and
		$\kappa = 0.001$ meV (top right), 2.6 meV and $\kappa = 0.0$ meV (bottom left),
		and 2.6 meV and $\kappa = 0.001$ meV (bottom right).
		$N_\mathrm{e} = 2$, $E_\gamma = 1.3$ meV, $g_\gamma = 0.06$, $pq = 1$,
		corresponding to $E_c = \hbar\omega_c = 0.7145$ meV, $V_t/\hbar\omega_c = 0.5$, and $T = 1$ K.}
	\label{2e-Eg13-t-spectra}
\end{figure*}

In Fig.\ \ref{2e-Eg13-t-spectra} are displayed the Fourier power spectra for
the cases shown in Fig.\ \ref{2e-Ng-13} with and without tiny photon dissipation.
For $E_\mathrm{ext} = 2.7$ meV and no dissipation (upper left) we notice a fundamental $N_\gamma(\omega)$
2-photon diamagnetic peak and its second harmonic. In addition we find a strong $Q_0(\omega)$ peak close to
$\omega/\omega_c\approx 11.5$. The addition of a tiny dissipation does not significantly change much
the diamagnetic peaks, but introduces paramagnetic peaks associated with 1-photon processes.

For $E_\mathrm{ext} = 2.6$ meV the peak heights are reduced and the strongest diamagnetic peaks are split
as was seen for the case of $N_\mathrm{e} = 1$ in Fig.\ \ref{1e-Eg13-t-spectra}. Clearly, in this case
the last large peaks of $Q_0(\omega)$ are reduced.

\begin{figure}[htb]
	\includegraphics[width=0.48\textwidth]{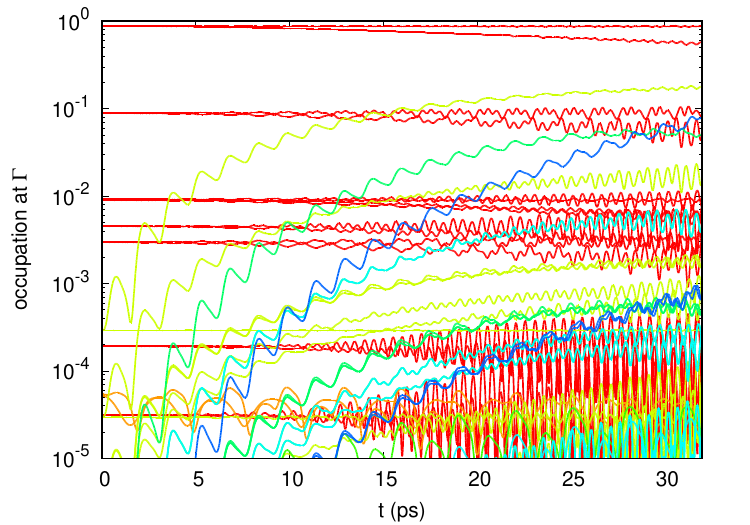}
	\includegraphics[width=0.48\textwidth]{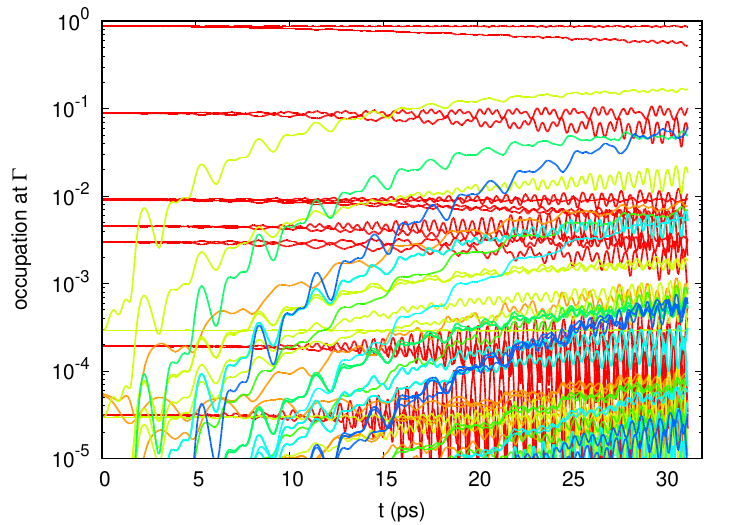}
	\caption{The occupation of the states $|\bm{\alpha\theta}n\sigma\rangle$ versus
		time for (top) $E_\mathrm{ext} = 2.7$ meV and $\kappa = 0.0$ meV,
		and (bottom) $E_\mathrm{ext} = 2.7$ meV and $\kappa = 0.001$ meV.
		The photon content of the states is color coded according
		to the scale used in Fig.\ \ref{1e-spectra}.
		$N_\mathrm{e} = 2$, $E_\gamma = 1.3$ meV, $g_\gamma = 0.06$, $pq = 1$,
		corresponding to $E_c = \hbar\omega_c = 0.7145$ meV, $V_t/\hbar\omega_c = 0.5$, and $T = 1$ K.}
	\label{2e-occ-dis}
\end{figure}

The occupation of the states close to the $\Gamma$ point in the reciprocal space for
$E_\mathrm{ext} = 2.7$ meV with (lower panel) and without (upper panel) photon dissipation
is presented in Fig.\ \ref{2e-occ-dis} and can be used to gain insight.
First, we note how the dissipation increases the occupation of a 1-photon state around the
center of the logarithmic scale in the lower panel. This state remains with a much lower occupation
without dissipation (upper panel). This is an example of a weak 1-photon paramagnetic transition from
a 2-photon state that was occupied through diamagnetic 2-photon excitation.

Importantly, we notice strong oscillation of 0-photon electron states (red) indicating intradot
radial or breathing oscillations. Clearly, they also lead to weaker corresponding interdot collective
oscillations. Close to the minima of the periodic potential the effective confinement energy of the
quantum dots can be estimated to be approximately 6 meV. The lowest breathing (or radial monopole) charge
oscillations in parabolically confined quantum dots has been calculated numerically and analytically estimated
to happen close to double the confinement frequency \cite{Abraham_2014,ANDP:ANDP201400048}.
Here, this estimate would give an energy of about $12$\,meV, corresponding to
$\omega/\omega_c \approx 16.8$. However, the parabolic approximation of the
minima in the periodic potential is not adequate for two electrons in a dot at
the low magnetic field corresponding to $pq=1$. We therefore realize that the strong $Q_0(\omega)$ peaks close to $\omega/\omega_c\approx 11.5$ arise from a collective breathing Coulomb mode located at a lower
frequency due to the tapering-off confinement potential away from its center.

\subsection{Li-like array ($N_\mathrm{e} = 3$)}
\label{Ne3-Results}
\begin{figure}[htb]
	\includegraphics[width=0.48\textwidth]{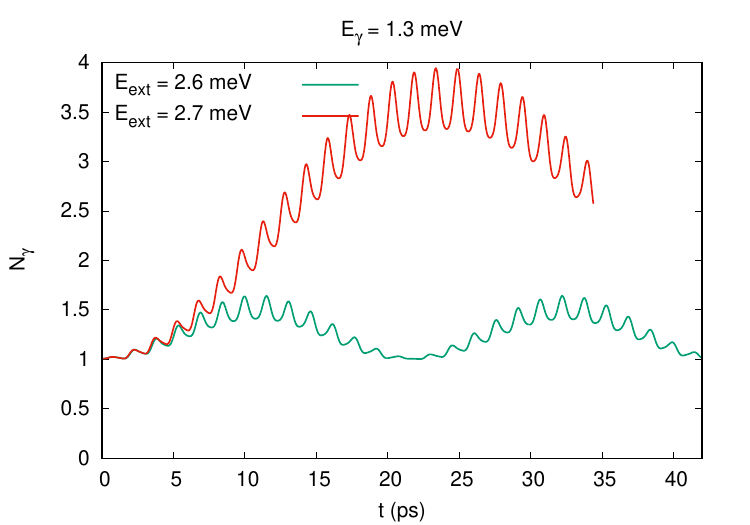}
	\caption{The mean photon number as function of time for two values of
		$E_\mathrm{ext}$ and $N_\mathrm{e} = 3$. $E_\gamma = 1.3$ meV, $g_\gamma = 0.06$, $pq = 1$,
		corresponding to $E_c = \hbar\omega_c = 0.7145$ meV, $V_t/\hbar\omega_c = 0.5$, and $T = 1$ K.}
	\label{3e-Ng-13}
\end{figure}

Fig.\ \ref{3e-Ng-13} presents the mean photon number per dot $N_\gamma(t)$ for
photon energy $E_\gamma = 1.3$ meV for $E_\mathrm{ext} = 2.6$ and 2.7 meV.
Here, the situation is different in the sense that for $N_\mathrm{e} = 3$ and the
photon energy 1.3 meV the lowest lying photon replica is occupied, as was discussed
in connection with the energy bandstructure shown for this case in the right panel
of Fig.\ \ref{23e-spectra}.

\begin{figure*}[htb]
	\includegraphics[width=0.48\textwidth]{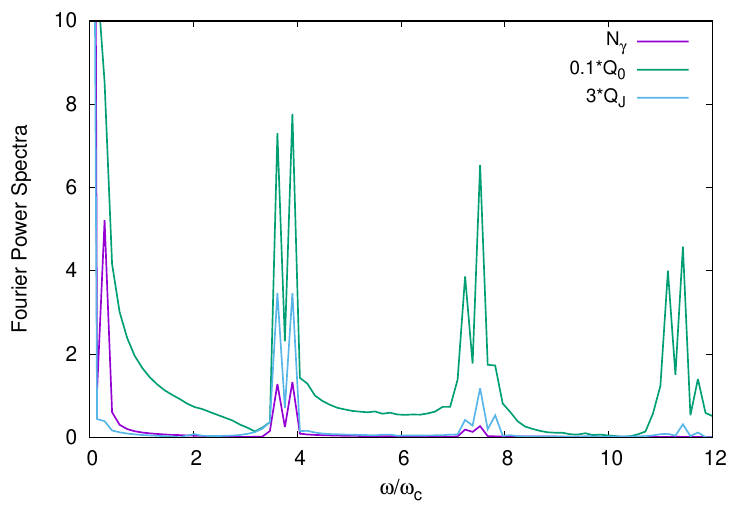}
	\includegraphics[width=0.48\textwidth]{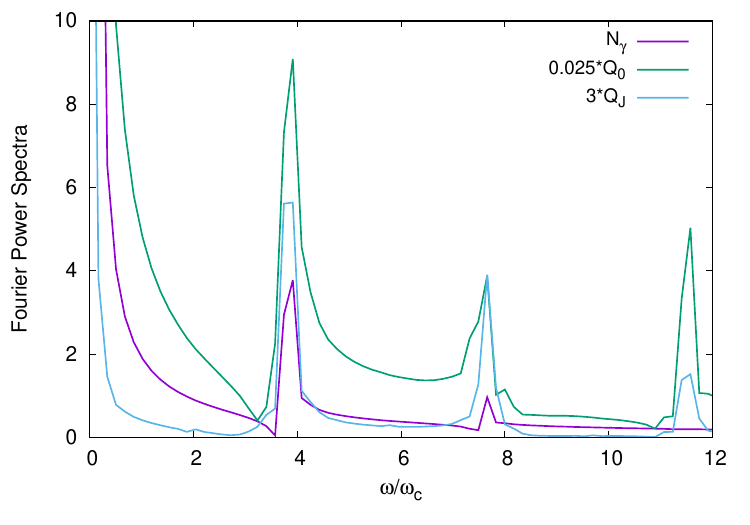}
	\includegraphics[width=0.48\textwidth]{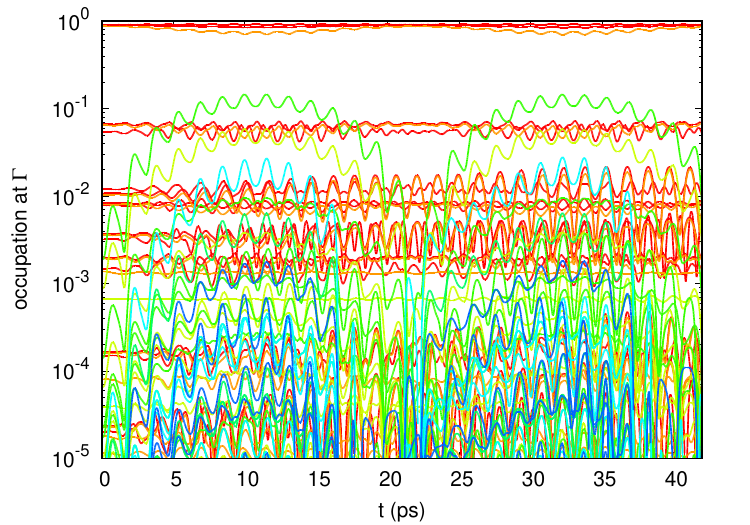}
	\includegraphics[width=0.48\textwidth]{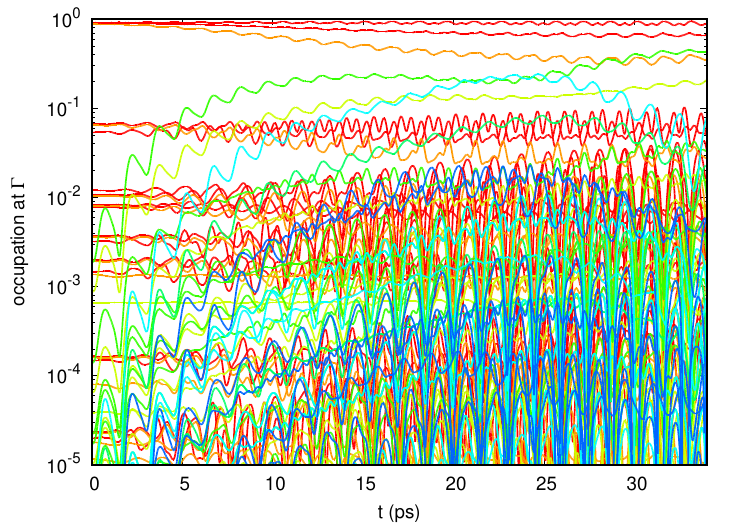}
	\caption{The Fourier power spectra $N_\gamma (\omega)$, $Q_0(\omega)$, and
		$Q_J(\omega)$ (top row) with $E_\mathrm{ext} = \hbar\omega_\mathrm{ext} = 2.6$ meV
		(left), and  $E_\mathrm{ext} = 2.7$ meV (right). In the lower row,
		the time-dependent occupation of the states $|\bm{\alpha\theta}n\sigma\rangle$ for
		$E_\mathrm{ext} = 2.6$ meV (left), and $E_\mathrm{ext} = 2.7$ meV (right).
		$N_\mathrm{e} = 3$, $E_\gamma = 1.3$ meV, $g_\gamma = 0.06$, $pq = 1$,
		corresponding to $E_c = \hbar\omega_c = 0.7145$ meV, $V_t/\hbar\omega_c = 0.5$, and $T = 1$ K.}
	\label{3e-13-rof-occ}
\end{figure*}

As expected, the photon number $N_\gamma(t)$ is close to a driving or pumping resonance
for $E_\mathrm{ext} = 2.7$ meV and strong peak caused by diamagnetic 2-photon processes is
seen in the upper right panel of Fig.\ \ref{3e-13-rof-occ}, and in the upper left panel of the figure for
$E_\mathrm{ext} = 2.6$ meV we again encounter split peaks. Just like for the $N_\mathrm{e} = 2$
case above in Fig.\ \ref{2e-Eg13-t-spectra} we notice strong $Q_0$ peaks close to
$\omega/\omega_c\approx 11.5$ representing interdot Coulomb radial breathing collective oscillations.

This last statement is supported by the corresponding occupancies close to the $\Gamma$ point
in the reciprocal space presented in the lower panels of Fig.\ \ref{3e-13-rof-occ}.
There the the high occupancy of the first photon replica (orange) is clearly seen together with
oscillating 0-photon and other states (red). As could be expected low order photon replicas become
occupied and higher order ones gain a larger, though small, occupation for $E_\mathrm{ext} = 2.7$ meV (right),
than for $E_\mathrm{ext} = 2.6$ meV (left).

Generally, we can claim that the number of photons pumped into each dot in the array grows
with the number of electrons in each dot. That could be expected as the electron charge is
fundamental to the electron-photon interaction.
\begin{figure}[htb]
	\includegraphics[width=0.48\textwidth]{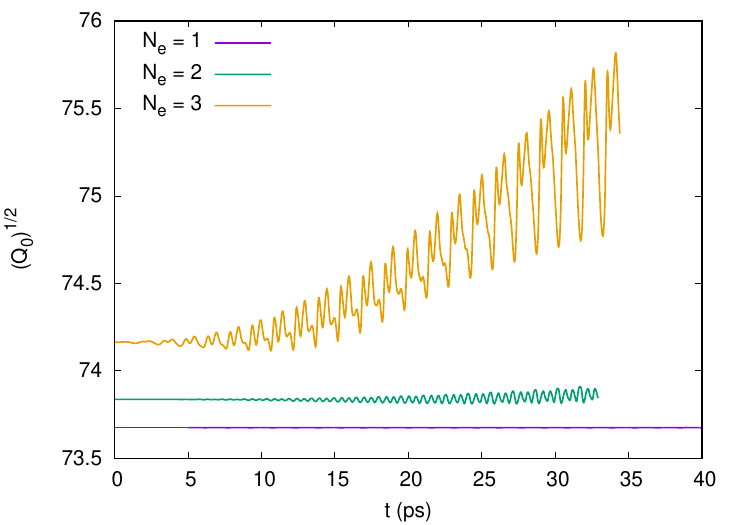}
	\includegraphics[width=0.48\textwidth]{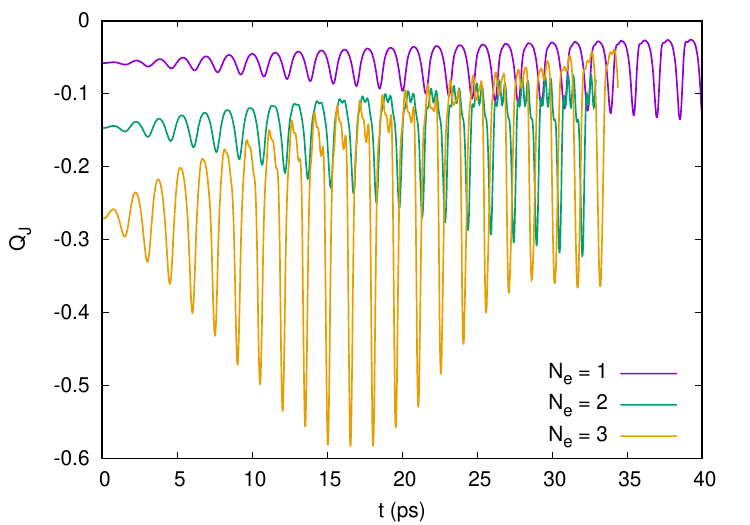}
	\caption{The mean diameter of the electron distribution of a quantum dot
	         $\sqrt{Q_0}$ (top), and the orbital angular momentum, or the dynamic orbital magnetization,
	         (bottom) for 3 different values of $N_\mathrm{e}$.
	         $E_\gamma = 1.3$ meV, $E_\mathrm{ext} = 2.7$ meV, $g_\gamma = 0.06$, $pq = 1$,
	         corresponding to $E_c = \hbar\omega_c = 0.7145$ meV, $V_t/\hbar\omega_c = 0.5$, and $T = 1$ K.}
	\label{123e-QJ-Q0}
\end{figure}
Finer details are though expected to depend on the shell-structure of the dots and how that
can be influenced by the cavity photon energy and electron-photon coupling strength. As an
example, the effective dynamic diameter of the dot structures investigated here is presented
in Fig.\ \ref{123e-QJ-Q0}. It is seen that the charge polarizability is highest for the
Li-like system with $N_\mathrm{e} = 3$, but the H-like one-electron system has the lowest charge
polarizability and the He-like system lies in between. Similar differences can be seen for the
dynamic orbital magnetization $Q_J$ seen in the lower panel of Fig.\ \ref{123e-QJ-Q0}.

\section{Summary and Conclusions}
\label{Conclusions}

We have presented and computationally analyzed a method for the continuous excitation of quantum dot arrays in a square lattice configuration, subject to an external homogeneous magnetic field and embedded in a cylindrical FIR photon cavity. The excitation is achieved through harmonic modulation of the electron-photon interaction, with the Coulomb interactions described within spin-density functional theory and the electron-photon coupling—including both paramagnetic and diamagnetic contributions—treated using a many-body configuration interaction approach. Our results demonstrate that the effectiveness of this driving mechanism depends strongly on the modulation frequency and the number of electrons per dot, with the cylindrical TE$_{011}$ cavity mode playing a crucial role in favoring diamagnetic electron-photon transitions.

Despite significant differences in the electronic structure arising from exchange forces—particularly evident in the H-like ($N_e = 1$), He-like ($N_e = 2$), and Li-like ($N_e = 3$) configurations—the system exhibits a universal resonance condition: maximum photon accumulation occurs when the modulation frequency is approximately twice the cavity photon frequency ($\hbar \omega_{\text{ext}} \approx 2E_\gamma$). At this resonance, diamagnetic two-photon transitions dominate, leading to efficient photon pumping, progressive occupation of higher-order photon replicas, and sustained energy influx into the system. Off-resonance driving produces oscillatory behavior in the photon number with minimal net energy gain, analogous to classical driven oscillator dynamics but enriched by the quantum many-body structure. For lower modulation frequencies or weaker effective electron-photon coupling, paramagnetic single-photon transitions can emerge and compete with the diamagnetic processes.

The continuous driving mechanism excites not only photonic degrees of freedom but also collective electronic excitations, particularly radial monopole (breathing) charge oscillations in systems with $N_e = 2$ and $3$. These collective modes, occurring at frequencies determined by the effective confinement potential modified by Coulomb interactions, couple to rotational charge currents through the Lorentz force. This coupling is mediated by the radial dependence of both the external magnetic field and the TE$_{011}$ cavity mode vector potential, thereby establishing a connection between the para- and diamagnetic parts of the electron-photon interaction. The photon pumping efficiency scales with the number of electrons per dot, with the Li-like three-electron system exhibiting the highest charge polarizability and photon absorption, reflecting the importance of shell structure and the effective electron-photon coupling strength in determining the system's response.

These findings establish cylindrical cavity photon modes as a versatile platform for controlling quantum dot array excitations through selective activation of diamagnetic transitions and collective Coulomb modes. The radial structure of the TE$_{011}$ mode suggests potential applications in generating structured vector beams in the far-infrared regime. More broadly, our work demonstrates that the interplay among harmonic modulation, photonic excitations, magnetic confinement, and many-body electron correlations in dimensionally reduced nanostructures can be exploited to engineer quantum states, with implications for quantum information processing and terahertz spectroscopy.

\begin{acknowledgments}
This work was financially supported by the Research
Fund of the University of Iceland Grant No.\ 92199, and
the Icelandic Infrastructure Fund for ``Icelandic Research
e-Infrastructure (IREI)''. The computations were performed
on resources provided by the Icelandic High Performance
Computing Center at the University of Iceland.
Mughnetsyan and V.G.\ acknowledge support by the Higher
Education and Science Committee of Armenia Grant No.\
24LCG-1C004). Mughnetsyan acknowledges support by the Higher
Education and Science Committee of Armenia Grant No.\ 24WS-1C040.
V.\ Moldoveanu acknowledges financial support from the Core Program of the National
Institute of Materials Physics, granted by the Romanian Ministry of Research, Innovation and
Digitalization under the Project No.\ PC2-PN23080202.
W.-H.K.\ acknowledges the support from the National Science and Technology Council, Taiwan, under
Grants No.\ NSTC 115-2918-I-845-003 and NSTC 114-2221-E-845-002, and is thankful for the hospitality
of the Science Institute of the University of Iceland.
H.-S.\ Goan acknowledges support from the National Science and Technology Council (NSTC), Taiwan, under Grants No.\ NSTC 113-2112-M-002-022-MY3,
No.\ NSTC 113-2119M-002-021, No.\ NSTC 114-2119-M-002-018, No.\ NSTC 114-2119-M-002-017-MY3, and NSTC 115-2119-M-002-005,
the support of Taiwan Semiconductor Research Institute (TSRI) through the Joint Developed Project (JDP), and from the Physics Division,
National Center for Theoretical Sciences (NCTS), Taiwan. H.-S.\ Goan also acknowledges support from the National Taiwan University under Grants
No.\ NTU-CC-115L8937, No.\ NTU-CC-115L893704 and No.\ NTU-CC-115L8512, as well as the support from the {\lq\lq}Center for Advanced Computing and Imaging
in Biomedicine (NTU-115L900702){\rq\rq} through the Featured Areas Research Center Program within the framework of
the Higher Education Sprout Project by the Ministry of Education (MOE), Taiwan.
J.-D.C.\ acknowledges the support from the National Science and Technology Council, Taiwan, under
Grants No.\ NSTC 115-2112-M-002-011 and No.\ NSTC 114-2112-M-002-033. J.-D.C.\ is also grateful
for the support from the Physics Division, National Center for Theoretical Sciences, Taiwan.
C.-S.T.\ acknowledges support from National United University under research contract No.\ 115-NUUPRJ-08.
\end{acknowledgments}

%----------------------------------------------------------------------------------------
%

\appendix
\section{Technical details for the numerical results}
\label{Tech-details}
In the static part of the calculations the constancy of the electron number in the unit cell
$N_\mathrm{e}$ is guaranteed by updating the chemical potential at each self-consistency iteration
of the DFT formalism. In the dynamic calculations the initial value of the density matrix and the
the symmetries of the L-vN equation take over this role.

As is seen in some figures, the time-dependent calculations are not all extended to 100 ps.
This is due to the restriction of 1 week wall-time for each run on the computational facilities
(IREI) of the University of Iceland. The time shown in the figures thus reflects the fact that
close to resonance conditions more iterations are needed for the self-consistency of the L-vN equation.

We use the 8 lowest Landau subbands, the 10 lowest eigenstates of the photon number operator,
and a $20\times 20$ nonequispaced grid in the first Brillouin zone of the reciprocal space, $\bm{\theta}$,
built on a repeated 4-point Gaussian quadrature. $pq=1$ is the number of magnetic flux units through
the unit cell of the periodic superlattice. A $13\times 13$ reciprocal lattice is
used, i.e.\ $G_i\in \{-6,\cdots,0,\cdots,6\}$ for $i=1\; \mbox{and}\; 2$.

The ground state, or static, and the dynamical calculations are done in a linear
functional basis constructed as a
tensor product (TP) of electron and photon states
$|\bm{\alpha\theta}\sigma n\rangle = |\bm{\alpha\theta}\sigma\rangle\otimes|n\rangle$.
The electron states were proposed by Ferrari \cite{Ferrari90:4598}, and used by
Gudmundsson \cite{Gudmundsson95:16744} and Silberbauer \cite{Silberbauer92:7355}.
The photon states are the eigenstates of the photon number operator
$N_\gamma = a^\dagger_\gamma a_\gamma$ with eigenvalue $n$. $\sigma$ is the spin label
$\{\uparrow\downarrow\}$, and $\bm{\alpha}$ is a composite quantum number for the Landau subband number.

{
The basis states proposed by Ferrari using the symmetric gauge, which is consistent with the
Coulomb gauge, are particularly convenient for numerical calculations in a square superlattice
as the Cartesian coordinates $x$ and $y$ are treated on equal footing, opposite to what happens
in the commonly used Landau gauge \cite{Ferrari90:4598,Pfannkuche92:12606,Silberbauer92:7355}.
}

The self-consistent diagonalization of the static total Hamiltonian leads to the states
$|\bm{\alpha\theta}\sigma)$, but the Liouville-von Neumann Eq.\ (\ref{L-vN}) is solved in
the $\{|\bm{\alpha\theta}\sigma n\rangle\}$-basis as most matrix elements are known analytically there
\cite{Gudmundsson95:16744,PhysRevB.110.205301}.

{All data used for preparation of figures can be requested from the corresponding author.}

%----------------------------------------------------------------------------------------
%

%\bibliography{mod_qd}
%apsrev4-2.bst 2019-01-14 (MD) hand-edited version of apsrev4-1.bst
%Control: key (0)
%Control: author (8) initials jnrlst
%Control: editor formatted (1) identically to author
%Control: production of article title (0) allowed
%Control: page (0) single
%Control: year (1) truncated
%Control: production of eprint (0) enabled
%

%--------------------------------------------------------------------------------------
\end{document}